\documentclass[fleqn,usenatbib]{mnras}

\usepackage[T1]{fontenc}

\DeclareRobustCommand{\VAN}[3]{#2}
\let\VANthebibliography\thebibliography
\def\thebibliography{\DeclareRobustCommand{\VAN}[3]{##3}\VANthebibliography}

\usepackage{ulem}
\usepackage{graphicx}	
\usepackage{amsmath}	
\usepackage{amssymb}	

\title[Star formation in closely interacting galaxies]{A FUV and optical study of star formation in closely interacting galaxies: star forming rings, tidal arms and nuclear outflows}

\author[Yadav et al.]{
Jyoti Yadav,$^{1,2}$\thanks{E-mail: jyoti@iiap.res.in}
Mousumi Das,$^{1}$
Sudhanshu Barway$^{1}$
Francoise Combes$^{3}$
\\
$^{1}$Indian Institute of Astrophysics, Koramangala II Block, Bangalore 560034, India\\
$^{2}$Pondicherry University, R.V. Nagar, Kalapet, 605014, Puducherry, India \\
$^{3}$Observatoire de Paris, LERMA, College de France, CNRS, PSL University, Sorbonne University, F-75014 Paris, France
}

\date{Accepted XXX. Received YYY; in original form ZZZ}

\pubyear{2015}

\begin{document}
\label{firstpage}
\pagerange{\pageref{firstpage}--\pageref{lastpage}}
\maketitle

\begin{abstract}
We present a study of the morphology of star formation and the associated nuclear activity in a sample of 8 closely interacting southern galaxies, which are in different stages of interaction, starting with nearly merged nuclei that have one prominent bulge to more widely spaced interacting galaxies. We have used Far-Ultraviolet (FUV) observations from the Ultraviolet Imaging telescope (UVIT), near-Infrared observations from the infrared survey facility telescope (IRSF) and archival optical data from the VLT/MUSE integral field spectrograph. 
Analysing resolved stellar populations across the disk of the interacting galaxies can provide unique insights into how interactions affect galaxy properties, such as morphology, star formation rates and chemical composition. We take advantage of the unprecedented capabilities of MUSE and UVIT to carry out a highly detailed spatially and spectrally resolved study of star formation rate, star formation histories, metallicity and AGN activity in the sample of eight interacting galaxies which are in different stages of interaction.
Most of our sample galaxies are gas-rich and show evidence of recent, massive star formation in tidal tails, rings and spiral arms. This is evident from their FUV and H$\alpha$ emissions, which trace young, massive star-forming regions. We compared the star formation rate in the barred and unbarred galaxies in our sample and found that the barred galaxies do not show significant enhancement in star formation rate or large-scale difference in star formation morphology compared to unbarred galaxies. IC5250 and NGC7733N, show extended nuclear outflows of size $\sim$ 5 kpc and 8 kpc respectively.
\end{abstract}




\section{Introduction}
Galaxy interactions are the main drivers of galaxy evolution. According to $\Lambda$ CDM,  these interactions were more common at higher redshifts when major mergers led to the formation of supermassive black holes and massive bulges \citep{Yee1995ApJ...445...37Y, Kartaltepe2007ApJS..172..320K}. Interacting galaxies in our low redshift Universe can help us study the role of mergers in the formation of massive galaxies \citep{bundy.etal.2009}. Simulations show that galaxies grow by accreting smaller galaxies \citep{rodriguez-gomez.etal.2016}. Galaxy collisions in major mergers are expected to be the most violent.
Galaxy interactions and mergers can modify the properties of galaxies such as their gas and stellar masses, chemical content, morphology, kinematic and dynamical properties \citep{Arp1966apg..book.....A, Sanders1996ARA&A..34..749S, Struck1999PhR...321....1S, Saintonge2012ApJ...758...73S,Duc2013LNP...861..327D, Ballesteros2015A&A...579A..45B}. 

The morphology of the interacting galaxies depends on multiple factors, such as their relative orbits, gas fraction, mass ratio, and initial morphology. Galaxy encounters lead to the redistribution of mass and angular momentum. Interactions/mergers can lead to bridges of hot and cold gas, plumes and tidal tails, which are composed of material pulled out from the galaxy's outer disks. Galaxy collisions can also result in accretion rings, polar rings and collisional rings \citep{ Lynds1976ApJ...209..382L, Schweizer1987nngp.proc...18S, Buta1996FCPh...17...95B}. The tidal tails, rings or bridges can host star formation. The debris in the tails can condense and form gravitationally bound tidal dwarf galaxies, which have baryonic masses in the same range as dwarf galaxies \citep{Barnes1992Natur.360..715B, Lelli2015A&A...584A.113L}. The observability of these tidal signatures depends on the viewing angle, redshift, wavelength of the observations, and depth of data \citep{Lotz2011ApJ...742..103L, Blumenthal2020MNRAS.492.2075B}.

In the local universe, star formation activity in galaxies is often associated with flyby interactions or galaxy mergers. Observational studies show that encounters between close pairs can lead to gas compression, which leads to starburst activity \citep{Braine1992A&A...264..433B, Yee1995ApJ...445...37Y, Tonnesen2012MNRAS.425.2313T}.
 Simulations have also shown that interactions can enhance the cold gas in galaxies, which can augment the star formation rate \citep{Moreno2021MNRAS.503.3113M}.
The mergers or flyby interactions can also lead to tidally induced spiral arms and slow rotating bars in galaxies \citep{Gerin1990A&A...230...37G,Salo1991A&A...243..118S,Miwa1998ApJ...499..149M}. The tidal torques during galaxy interactions can lead to gas inflow into the central regions, leading to nuclear starburst and active galactic nucleus (AGN) activity \citep{ Barnes1992ARA&A..30..705B,Mihos1996ApJ...464..641M}. \citet{Knapen2015MNRAS.454.1742K} showed that interactions/mergers could enhance median star formation by a factor of 1.9. Using ultraviolet (UV) photometry from \textit{GALEX}, \citet{Wong2011ApJ...728..119W} showed that galaxies with projected separation $\leq$ 50 h$^{-1}$ kpc show an increase in specific SFR by $\sim$ 15-20 \% while the galaxies with projected separation $\leq $30 h$^{-1}$ kpc shows $\sim$ 25–30 per cent increase in sSFR. Studies using TNG simulations and SDSS data have also shown that tidal forces due to the interacting galaxies having projected separation as high as 150 kpc can lead to enhancement in star formation rate \citep{Patton2020MNRAS.494.4969P}.

\citet{Hutchings1987ApJ...320..122H} showed that AGN activity is also associated with mergers.
Ultraluminous infrared galaxies also show higher AGN fraction \citep{Kartaltepe2010}. \citet{Woods2010AJ....139.1857W} showed that bursts of star formation are observed during gas-rich galaxy interaction, as there is abundant fuel for new stars. However, gas-poor mergers do not have fuel for star formation, so they do not show starburst activity \citep{Bell2006ApJ...640..241B,Naab2006ApJ...636L..81N,Lin2008ApJ...681..232L}. The connection between merging galaxies and AGN is still debated. Multiple studies show significant AGN enhancement in merging galaxies \citep{Alonso2007MNRAS.375.1017A, Woods2007AJ....134..527W, Weston2017MNRAS.464.3882W} while other studies do not find a higher merger fraction in AGNs \citep{Grogin2005ApJ...627L..97G, Kocevski2012ApJ...744..148K}.

The investigation of stellar populations has been used as an approach in multiple studies to understand the evolution of interacting systems. Observational studies using integral field spectroscopy surveys such as CALIFA \citep{Ballesteros2015A&A...579A..45B}, MaNGA \citep{Steffen2023ApJ...942..107S}, and SAMI \citep{Bloom2018MNRAS.476.2339B}
have revealed critical information about central star formation, metallicity, kinematic asymmetry and AGN activity in close pairs to merger remnants. The study of resolved stellar population morphology is limited by the resolution of these large surveys. The resolution of MUSE is better than these studies ($\sim$ 1\arcsec). Massive, young star-forming regions in interacting galaxies are bright in the H$\alpha$ and far-UV (FUV) bands. The H$\alpha$ maps collapsed from MUSE data allow us to detect star-forming complexes with much better resolution. The FUV band of the Ultraviolet Imaging Telescope (UVIT) has a resolution of 1\arcsec, allowing us to trace star-forming regions with a relatively high spatial resolution. Hence, MUSE data and UVIT observations allow us to study star formation over a wide range of wavelengths and star formation timescale. Furthermore, AGN activity in the galaxies can also be studied using UVIT and MUSE observations. The following sections present a study of star formation and AGN activity in a sample of 8 interacting galaxies in different interaction stages. Using this sample, we examine how interactions, bars, and AGN activity affects star formation in galaxies. We have adopted a flat cosmology with $\Omega_{\Lambda}$=0.7, $\Omega_{M}$=0.3, and H$_{0}$= 70 km s$^{-1}$ Mpc$^{-1}$.

\begin{figure*}
    \centering
    \includegraphics[width=1\textwidth]{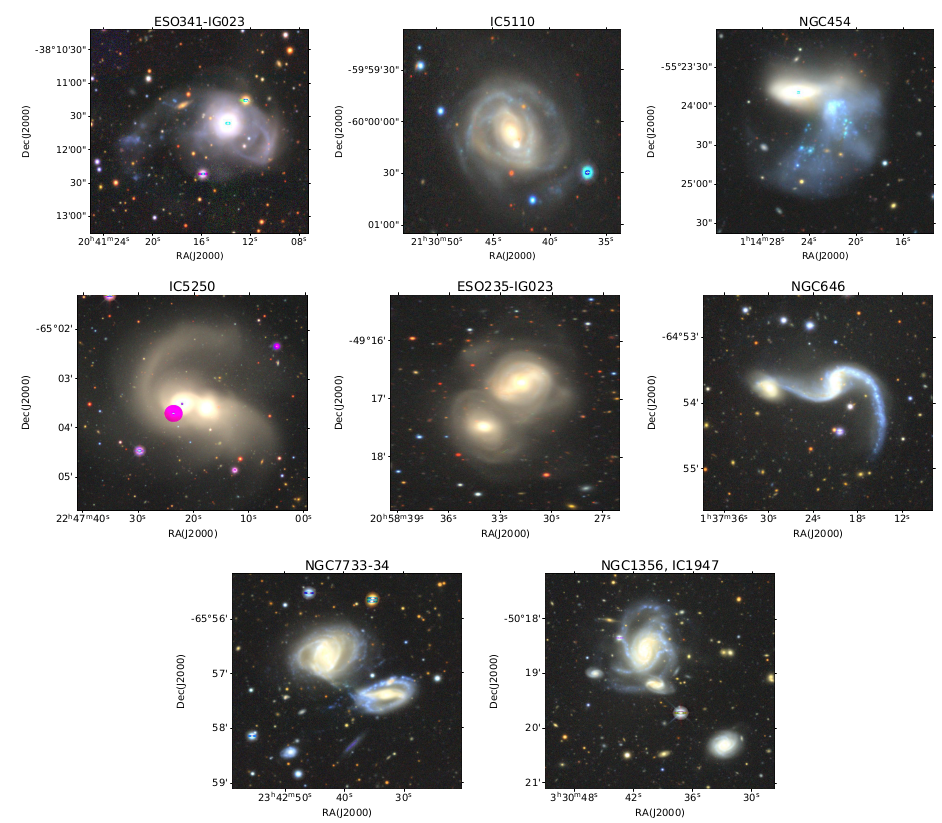} 
    \caption{Colour composite images of the sample galaxies in grz band from Dark Energy Camera Legacy Survey (DECaLS) (DR9 and DR10) \citep{decals92021AAS...23723503S}.}
    \label{fig:decals}
\end{figure*}

\begin{figure*}
    \centering
\includegraphics[width=0.75\textwidth]{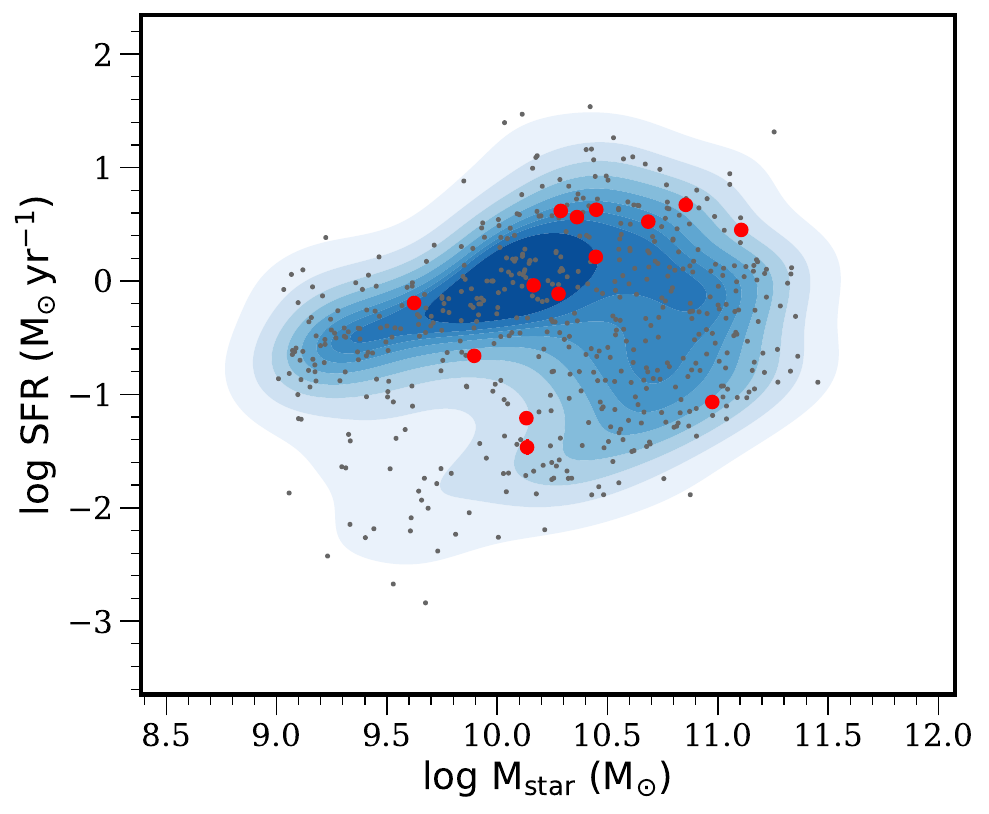}
    \caption{The star-forming main sequence of galaxies. The red symbols represent the galaxies in the sample. The SFR for the sample galaxies shown is derived using FUV data. The black symbols represent the distribution of xCOLD GASS sample \citet{Saintonge2017ApJS..233...22S}. The grayscale contour shows the density of the data points in the xCOLD GASS sample. }
    \label{fig:sfms}
\end{figure*}

\begin{table*}
    \centering
    \begin{tabular}{lccccccc}
    \hline
Source  & R.A. & Dec. & Distance & Galaxy Morphology & M(HI) & M(H$_{2}$) & M$_{\star}$\\
& (hh mm ss) & (dd mm ss) & (Mpc) & & [10$^{9}$ M\textsubscript{\(\odot\)}] & [10$^{8}$ M\textsubscript{\(\odot\)}] &  [10$^{10}$ M\textsubscript{\(\odot\)}]  \\
  \hline
ESO341-IG004$^{\dagger}$ & 20 41 13.87 &-38 11 36.3 & 88.2  & S0+: pec   &  3.9  &  17.6  & 2.3\\
\hline
IC5110 & 21 30 43.39 & -60 00 06.6  & 129 & S... + compact comp   &  --  &  < 9  & 1.46\\
\hline
NGC454NED01 & 01:14:21.810 & -55:24:05.10 & 51.8 & 	Irr   &  < 0.5$^{*}$ &  -- & 1.94 \\
NGC454NED02 & 01:14:24.93 & -55:23:49.5 & 52 & 	Irr   &     &  < 2 & 0.42\\
\hline
IC5250A & 22 47 17.50 & -65 03 35.0 & 45.4 & S   &   < 0.4$^{*}$  & < 1.8 & 1.36 \\
IC5250B & 22 47 22.16 & -65 03 30.6 & 45.7 & S0  &    &  --  & 9.43\\
\hline
ESO235-IG023NED01$^{\dagger}$ & 20:58:31.82 & -49:16:43.2 & 96.5 & (R)SB(r)0/a pec   &  <1.9 $^{*}$  &  $\leqslant$ 5.8  & 2.8\\
ESO235-IG023NED02$^{\dagger}$ & 20:58:33.96 & -49:17:28.9 & -- & S0   &    & $\leqslant$11.6  & 1.37 \\
\hline
NGC646NED01 & 01:37:21.12 & -64:53:41.5 & 123 & SBc   & 30.6$^{*}$   &  61.9 &  2.81 \\
 NGC646NED02  & 01:37:29.89 & -64:53:46.6 & 108 & --   &    & --  & 0.786  \\ 	
\hline
 NGC7733 & 23 42 32.95  & -65 57 23.4 & 152  & (R')SB(rs)b pec    &  $\leqslant$ 4.2$^{*}$  &  $\leqslant$ 15.4  & 24.5 \\
NGC7734 & 23 42 42.93  & -65 56 40.7 & 158 & (R')SB(r)b pec:   &    &  $\simeq$ 107.3  & 64.6 \\
NGC7733N & 23:42:32.250 & -65:57:09.43 & 152.2 &   &  --  &   -- & 0.912  \\
\hline
 NGC1356 & 03 30 40.79  & -50 18 34.6  & 178  & SAB(r)bc pec:  &  < 6  &  142.3 & 7.15  \\
 IC1947 & 03 30 32.82 & -50 20 18.7 & 174 & S...  &  --  & --  & 1.94  \\
 \hline
    \end{tabular}
    \caption{Details of the galaxies in the sample. Columns 2 and 3 are the coordinates of galaxies in J2000.0. Column 4 is the distance in Mpc. Column 5 is the morphology of the galaxy. Columns 6 and 7 are the HI and H$_{2}$ masses of galaxies. Column 8 represents the stellar mass of galaxies estimated after GALFIT fitting. The coordinates, distance and galaxy morphology are listed from NED. The HI and H$_{2}$ masses are listed from \citet{Horellou1997A&AS..126....3H}. $*$It is the total mass in the system. $^{\dagger}$MUSE data is unavailable for two interacting systems, ESO341-IG004 and ESO235-IG023.}
    \label{tab:details}
\end{table*}
\subsection{Sample selection}
We have a sample of eight southern interacting galaxies in different stages of interaction, both merged systems and separated galaxies with tidal arms and bridges. The galaxy sample was drawn from a study of about 60 gas-rich interacting/merging galaxies in the southern sky \citep{Horellou1997A&AS..126....3H}. Since our aim was to study star formation morphology in merging galaxies, we selected gas-rich galaxies from this sample with star-forming complexes (SFCs) in their tidal tails and arms. The galaxies are in different stages of interaction; some of them are in the advanced stage of merging, e.g. ESO341-IG023, IC5110 and NGC454. Some show two separate galaxies interacting at a distance, e.g. IC5250, ESO235-IG023, NGC7733-33N-34, NGC1356 and IC1947. NGC646 appears to be a fly-by interaction, where the interaction has induced long tidal tails in NGC646NED01. These galaxies show plumes, loops, tidal tails, disturbed morphologies and merger remnants.
We carried out UVIT observations of these galaxies. The UVIT has high sensitivity and very good spatial resolution (200 pc to around 800 pc at the distance of the galaxies in the sample), which helps to isolate the star-forming knots, image the diffuse UV emission, and detect tidal dwarf galaxies. We also carried out near-infrared (NIR) observations using the infrared survey facility telescope (IRSF). The details of the galaxies are given in Table \ref{tab:details}.

\begin{table*}
    \centering
    \caption{FUV observation details of sample galaxies: column 1: Sample Galaxy; column 2: Proposal ID; column 3: name of the FUV filter, column 4: exposure time of FUV observation}
    \begin{tabular}{lccc}
    \hline
Source & Proposal ID & FUV Filter & Exp Time (sec) \\
\hline
ESO341-IG004 & A05-101 &  F148W (CaF2-1)  &  2569.85   \\
IC5110 &  A05-101  &    F148W (CaF2-1)   & 2850.02  \\
NGC454 &    AIS-392-1-44    &   \textit{GALEX} (FUV)  &   189     \\
IC5250 &   T04-095  &   F148W (CaF2-1)   & 6959.767   \\
ESO235-IG023 & A05-101 &  F148W (CaF2-1) &  2875.326  \\
NGC646 &  A05-101  &  F148W (CaF2-1)  & 2844.304  \\
NGC7733-NGC7734 & A05-101 &  F148W (CaF2-1)  &  2577.869  \\
NGC1356-IC1947 & GI1-047019-IC1959 &  \textit{GALEX} (FUV) &  1508.05   \\   
    \hline
    \end{tabular}
    \label{tab:observations}
\end{table*}

\section{Observations}
\subsection{FUV Data}
We observed a sample of eight southern interacting galaxies in FUV using UVIT onboard The \textit{Astrosat} Satellite \citep{2012Kumar}. The UVIT has a spatial resolution of $\approx$1.2\arcsec and a field of view of 28\arcmin. The UVIT can observe in FUV (1300-1800 \AA) and near UV (NUV) (2000-3000 \AA) and has multiple photometric filters. A visible channel (3200-5500 \AA) was used for the drift correction. \footnote{\url{https://astrobrowse.issdc.gov.in/astro_archive/archive/Home.jsp}}. We used FUV data for ESO341-IG004, IC5110, IC5250, ESO235-IG023, NGC646, NGC7733, and NGC7734, from UVIT.
We reduced UVIT Level 1 data using CCDLAB \citep{Joe2017}. We corrected for flat fields, cosmic rays, field distortions and drift. UVIT data was unavailable for NGC454 and NGC1356 due to payload-related issues. Hence, we used \textit{GALEX} data for these galaxies. \textit{GALEX} has a field of view of 1.25 $^{\circ}$. \textit{GALEX} has two filters, FUV (1344-1786 \AA)  and NUV (1771-2831 \AA) \citep{Morrissey2007}. \textit{GALEX} has an angular resolution of 4.2 \arcsec and 5.3 \arcsec in FUV and NUV, respectively. The details of the FUV observations are given in Table \ref{tab:observations}.

\subsection{Near Infrared Data}
We also observed the sample galaxies in the NIR using the SIRIUS camera \citep{2003SPIE.4841..459N} on the Infrared Survey Facility (IRSF) 1.4 m telescope in Sutherland, South Africa. The telescope can simultaneously observe in the J, H, and K bands. The IRSF camera has a pixel scale of 0.45\arcsec\ and a field of view of 7.7\arcmin\ $\times$ 7.7\arcmin\ \citep{2012Nagayama}. The images have a total exposure time of $\sim$120 minutes and seeing of $\sim$ 1.5\arcsec. The images were taken in the dithering mode with an exposure time of 30 seconds each. We reduced the data using a pipeline for the SIRIUS observations. We corrected for dark subtraction, flat fielding and non-linearity effects. It is well known that NIR imaging is the best way to reveal stellar masses \citep{Zibetti2009MNRAS.400.1181Z, Nagaraj2021ApJ...912..145N}. This is especially important for interacting galaxies, where the morphology is often disturbed.

\subsection{MUSE Data}
To understand the H$\alpha$ distribution, we used archival data from the Multi-Unit Spectroscopic Explorer (MUSE). The MUSE data is provided in a reduced form from the European Southern Observatory archive. MUSE is an integral field spectrograph on the Very Large Telescope (VLT). It can give 3D spectroscopic data cubes with high spatial and spectral resolution. MUSE has a field of view of 1\arcmin\ $\times$ 1\arcmin, a spatial resolution of $\sim$ 0.3-0.4 \arcsec\ and a spectral resolution of 1750 at 4650{\AA} to 3750 at 9300\AA. The seeing of the sample galaxies vary from 1--2\arcsec.
We used the Galaxy IFU Spectroscopy Tool (GIST\footnote{\url{https://abittner.gitlab.io/thegistpipeline/}} version 3.0.3; \citealt{2019Bittner}) pipeline to analyse the MUSE data. GIST uses GANDALF \citep{2006Sarzi, 2006Falcon, 2019Bittner} for emission line and Penalized Pixel-Fitting (pPXF) \citep{2004Cappellari,2017Cappellari} method for stellar continuum fitting. GIST can bin the data for a given signal-to-noise (S/N) ratio over the provided wavelength range and also allows the user to remove noisy data below an S/N level before binning the data.

We have run the gist pipeline for stellar kinematics and gas emission line fitting. We binned the data based on H$\alpha$ emission using Voronoi binning. We used an S/N of 50 for binning the data and an S/N of 3 to remove the noisy data. We have also corrected the spectra for the Milky Way extinction. We fitted the continuum by multiplicative eight-order Legendre polynomial. The galaxies in the sample are star-forming and brighter in H$\alpha$, so the bins are smaller and finer.

\begin{table*}
    \centering
    \begin{tabular}{lccccc}
    \hline
  Source & No. of detected SFCs  &  & $\Sigma_{SFR(FUV)}$ & $\Sigma_{SFR(H\alpha)}$   & AGN Activity\\
 &  FUV  &  H$\alpha$ &(M$_\odot$ yr$^{-1}$ kpc$^{-2}$) & (M$_\odot$ yr$^{-1}$ kpc$^{-2}$) & \\
  \hline
  ESO341-IG004 & 5 & --& 0.2090$\pm$0.0251 &-- &  --  \\
  \hline
  IC5110 & 7 & 35 & 0.0068$\pm$0.0015 & 0.0616$\pm$0.0009 &  No \\ 
  \hline
  NGC454NED01 & 3 & 8 & 0.0914$\pm$0.0275 & 0.4266$\pm$0.0042  &  No \\ 
  NGC454NED02 & -- & -- & -- & -- & Yes \\  
  \hline
  IC5250A & -- & -- &  -- & --  & No  \\
  IC5250B & -- & -- &  -- & --  & Yes  \\
  \hline
  ESO235-IG023NED01 & -- & -- & -1.92$^*$ & -- &  -- \\
  ESO235-IG023NED02  & -- & -- & -- & --  &  --\\ 
  \hline
   NGC646NED01 & 25 & 22 & 0.0147$\pm$0.0008 & 0.4509$\pm$ 0.0046 &  Unclear  \\
   NGC646NED02  &  1 & 5  & 0.0223$\pm$0.0012  & 0.2597$\pm$0.0028 &  Unclear\\ 	
   \hline
  NGC7733 & 27  & 55 & 0.0081$\pm$0.0007 & 0.1946$\pm$0.0023  & Yes  \\
  NGC7734 & 12 & 41 & 0.0143$\pm$0.0033 & 0.2009$\pm$0.0028  &  Yes \\
  NGC7733N & -- & -- & -- &  -- & Yes\\
  \hline
  NGC1356 & 18  & 86 & 0.0580$\pm$0.0039 & 0.1309$\pm$0.0017 &  No \\
  IC1947 & 2 & 24 & 0.0821$\pm$0.0041 & 0.0727$\pm$0.0009 &   Yes \\
 \hline
    \end{tabular}
    \caption{Columns 2 and 3 are the number of detected SFCs in FUV and H$\alpha$. Columns 4 and 5 are the mean $\Sigma_{SFR}$ of galaxies in FUV and H$\alpha$.  Column 6 represents whether the galaxy hosts an AGN or not. $^*$ The $\Sigma_{SFR}$ of ESO235-IG023NED01 was calculated using an elliptical annulus around the ring.}
    \label{tab:sfcs}
\end{table*}

\section{Analysis}\label{sec:Analysis}

\subsection{Star Formation Main Sequence for the Sample Galaxies}

We used GALFIT \citep{Peng2002AJ....124..266P} to perform the 2D decomposition of galaxies in the IRSF J, H and K bands. We have fitted multiple sersic profiles to the galaxies to fit the bulge, disk, and bar components. The NIR luminosity from a galaxy is associated with its stellar content, and the relation between mass to light is nearly constant in the NIR band, as the young stellar population contribute less in the NIR. To calculate the stellar mass of the sample galaxies, we used the K band apparent magnitude obtained after the decomposition of individual components in galaxies. We corrected the magnitude for Galactic extinction using the calibration of  \citep{Schlafly2011ApJ...737..103S} and converted the magnitude to luminosity. We estimated the mass of the components in the galaxy using M/L= 0.6 \citep{McGaugh2014AJ....148...77M} and combined the masses of the individual components to obtain the total stellar mass. The stellar masses of the galaxies are given in Table \ref{tab:details}. We have not presented the detailed results from NIR analysis using GALFIT in this work. Detailed analysis of the NIR imaging of the sample will be presented in a future study (in prep.). We also estimated the integrated SFR using FUV data. The details about the SFR estimation are given in Sec. \ref{sec:sfr_estimation}. Fig. \ref{fig:sfms} shows the star formation main sequence of the sample galaxies. The red symbols represent the galaxies in the current sample, and the black symbols represent the galaxies in the xCOLD GASS sample from \citet{Saintonge2017ApJS..233...22S}

\subsection{SFCs detection and estimation of SFR}\label{sec:sfr_estimation}
The galaxies in the sample show a significant amount of star formation in the disk and the tidal tails. The star-forming regions are visible in FUV and H$\alpha$ emission as bright knots. We used the Python library for Source Extraction and Photometry (SExtractor, \citealt{Bertin1996}) to identify and extract the SFCs in the galaxies. SExtractor can identify the source based on a given threshold. The pixels with counts exceeding the threshold are identified as a source. We used the detection threshold to be 5$\sigma$, where $\sigma$ is the global background noise, i.e., the RMS of the counts. The extracted sources are referred to as the SFCs in this work.
We used Photutils, a Python astropy package for photometry, to perform elliptical aperture photometry on the detected sources. We have corrected foreground Galactic extinction using the Fitzpatrick law \citep{fitzpatrick1999} by assuming optical total to selective extinction ratio R(V)=3.1. 
\begin{equation}
A_{\lambda}  =R_{\lambda}\times E(B-V)
\end{equation}
where $A_{\lambda}$ is the extinction at wavelength $\lambda$, and E(B-V) is the reddening.
We corrected the internal extinction using Balmer decrement from MUSE data. For star-forming galaxies, Balmer decrement stays constant for typical gas conditions \citep{Osterbrock1989SvA....33..694O}. We consider the value of $(H_{\alpha}/H_{\beta})_{int}$ = 2.86 which corresponds to temperature of 10$^{4}$ K and electron density of 10$^{2}$ cm$^{-3}$ for Case B recombination \citep{Osterbrock1989SvA....33..694O}. The following equation gives the nebular colour excess:  
\begin{equation}
    E(B-V) = 1.97 \log_{10}\Biggr[\frac{(H_{\alpha}/H_{\beta})_{obs}}{2.86}\Biggr]
\end{equation}

\citet{Kennicutt1998ARA&A..36..189K} suggested that disk star-forming galaxies have H$\alpha$ extinction  A$_{\alpha}$= 1.1 mag and A$_{FUV}$/A$_{H\alpha}$= 3.6 if the FUV emission comes from young and massive star forming regions \citep{Leroy2012AJ....144....3L}. For the galaxies where MUSE data is unavailable, we corrected the internal extinction using A$_{\alpha}$= 1.1 mag.
We calculated the $\Sigma_{SFR}$ in FUV for each of the SFCs in our galaxies using the following formula \citep{Leroy2012AJ....144....3L, Salim2007}:
\begin{equation}
\label{eqn:sfruv}
SFR_{(FUV)}  = 0.68\times10^{-28} L_{FUV}
\end{equation} 
Where SFR$_{(FUV)}$ is the SFR in FUV [M$_\odot$ yr$^{-1}$] and $L_{FUV}$ is the FUV Luminosity [erg s$^{-1}$ Hz$^{-1}$].

We used \citet{Calzetti2007ApJ...666..870C} law to calculate the SFR for each SFC in H$\alpha$.
\begin{equation}
\label{eqn:sfrha}
SFR_{(H\alpha)}  = 5.3\times10^{-42} L_{H\alpha}
\end{equation} 
Where SFR$_{(H\alpha)}$ is the SFR in H$\alpha$ [M$_\odot$ yr$^{-1}$] and L$_{H\alpha}$ is the H$\alpha$ Luminosity [erg s$^{-1}$].

The number of identified SFCs in FUV and H$\alpha$ and their SFR/area ($\Sigma_{SFR}$) is given in Table \ref{tab:sfcs}.

\subsection{Estimation of SFH, metallicity and velocity}
For estimating the non-parametric star formation history (SFH), the SFH module in the GIST pipeline uses emission line subtracted spectra produced by the GANDALF module. The stellar population properties and non-parametric SFH is estimated using pPXF by modelling the observed spectrum. It produces a linear combination of spectral templates and ascribes them linear weights to match the model spectra with the observed spectra. The SFH and stellar population properties are estimated using the linear weights given to the template spectra. A regularization algorithm is used to dampen the high-frequency variation in the solution. Regularization provides the smoothest solution among the other degenerate solution, which can also describe the data. 

Metallicity measurements are important for understanding the evolutionary history of galaxies. The gas-phase metallicity gives us information about the instantaneous oxygen abundance. Oxygen is mainly supplied to the galaxy by core-collapse supernovae.
GIST uses the GANDALF module to derive gaseous kinematics and emission line properties. GANDALF module provides emission line flux, amplitude, velocity and amplitude-to-noise ratio. We used emission line fluxes obtained from GANDALF to derive the gas-phase metallicity map of each galaxy.
Metallicity maps are derived using \citet{Pettini2004MNRAS.348L..59P}.
\begin{equation}\label{eq:metallicity}
    12+log(O/H)= 8.73-0.32\times log[([\ion{O}{iii}]/H\beta)/ ([\ion{N}{ii}]/H\alpha)]
\end{equation}

Galaxy mergers also play a crucial role in changing the stellar and gas kinematics of galaxies. Multiple studies in the past have revealed surprising misalignment between the gas and stellar disks in galaxies. The tidal forces during galaxy interactions and mergers can perturb the stellar and gas disk. The gas stripping during galaxy interactions can also lead to misaligned gas and stellar disk \citep{Starkenburg2019ApJ...878..143S}. 
\citet{Barrera2015A&A...582A..21B} have also suggested that ionized gas reacts more easily to interaction than the stellar disk. The galaxies in the sample are star-forming and much brighter in H$\alpha$. We derived H$\alpha$ and stellar velocity using the GIST pipeline to check the perturbation in velocities due to interaction.

\subsection{BPT and WHAN diagram}
Studying the AGN and star formation activity in interacting galaxies is very important as it helps us understand the role of galaxy interactions in triggering star formation and feeding supermassive black holes. The Baldwin–Phillips–Terlevich (BPT, \citealt{1981Baldwin}) diagram classifies the source of the ionization mechanism of the gas based on the flux ratios of the emission lines.
\citet{2001Kewley} constructed a division line between the star-forming region and AGN by considering the population synthesis and photoionisation models. The SDSS survey showed that galaxies tend to fall into two sequences, resulting in the seagull shape. The left-wing shows ionization due to star-forming regions, and the right-wing shows ionization due to AGN. Later \citet{2003Kauffmann} proposed an empirical new distinction between star formation and AGN based on observation data of 10$^{5}$ galaxies from SDSS. The right wing of the BPT diagram shows bimodality with an upper and lower branch. The objects in the upper and lower branches are Seyfert and LINER AGN. To study the nature of the emission mechanism in our sample of galaxies, we derived their BPT plots using MUSE data cubes. The BPT diagram uses the [\ion{O}{iii}]$\lambda 5007/$H$\beta$ and the [\ion{N}{ii}]$\lambda 6563/$H$\alpha$ line ratios to distinguish between Seyfert, Liner, composite and star-forming regions.

\citet{2008Stasinska} showed that the emission from retired galaxies has the same location as LINERs in the BPT diagram. Retired galaxies have stopped forming stars. They have hot, low-mass-evolved stars such as hot post-AGB and white dwarfs, which can ionize the surrounding medium and produce line ratios similar to LINER.
They showed that a significant amount of LINERs in the BPT right seagull wing are retired galaxies.
\citet{Cid2011MNRAS.413.1687C} proposed a WHAN diagram, which considers the equivalent width of H$\alpha$ and can distinguish between emission from LINER and retired galaxy.

\begin{figure*}
    \centering
    \includegraphics[width=0.8\textwidth]{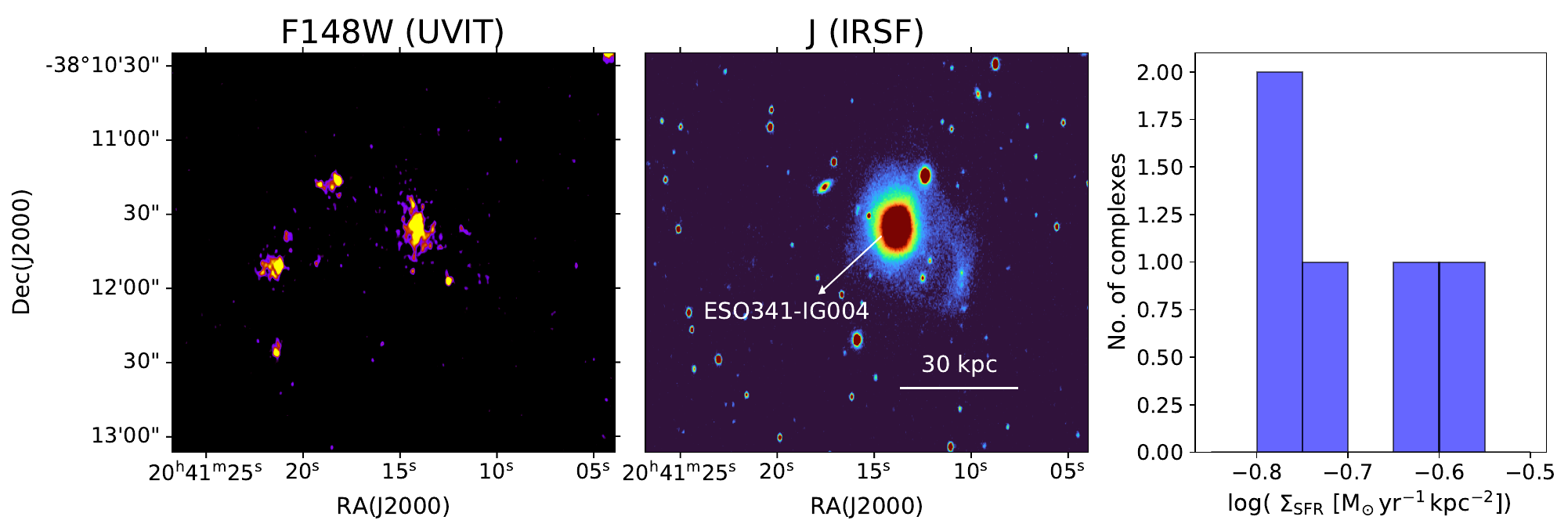}
    \caption{Left and the middle panel show the image of ESO341-IG004 in the FUV and J bands, respectively. The right panel shows the $\Sigma_{SFR(FUV)}$ of SFCs in ESO341-IG004 plotted as a histogram}.
    \label{fig:eso341}
\end{figure*}

\begin{figure*}
    \centering
    \includegraphics[width=0.8\textwidth]{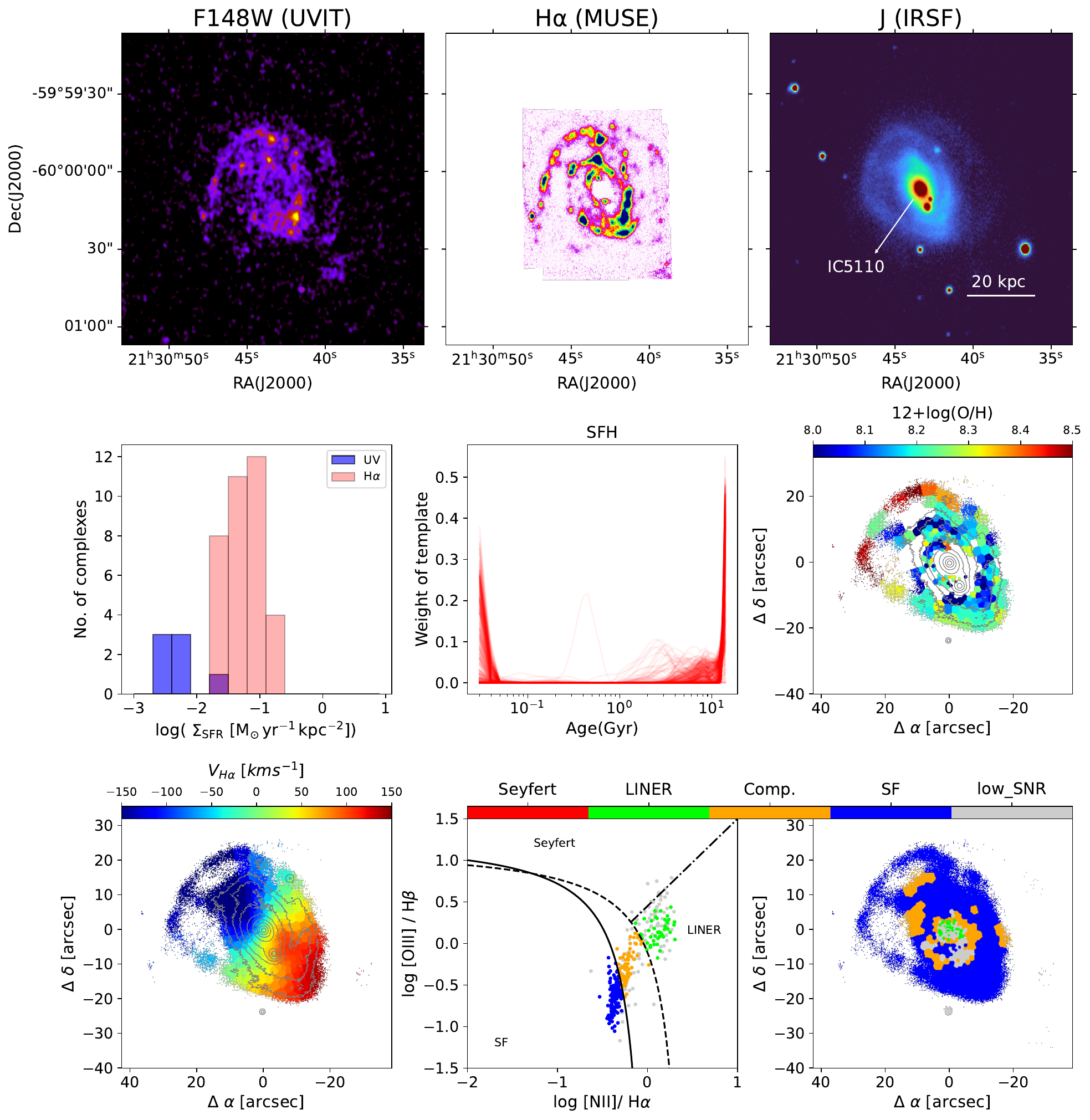}
    \caption{The top left, middle and right panels show the images of IC5110 in FUV, H$\alpha$ and J bands, respectively. The middle left panel shows  the histograms of $\Sigma_{SFR(FUV)}$ and $\Sigma_{SFR(H\alpha)}$. The middle and right panels show the SFH and metallicity. The bottom left panel shows the V$_{H_\alpha}$ velocity. The bottom right panel shows the BPT map of the galaxy.}
    \label{fig:IC5110}
\end{figure*}

\begin{figure*}
    \centering
    \includegraphics[width=0.8\textwidth]{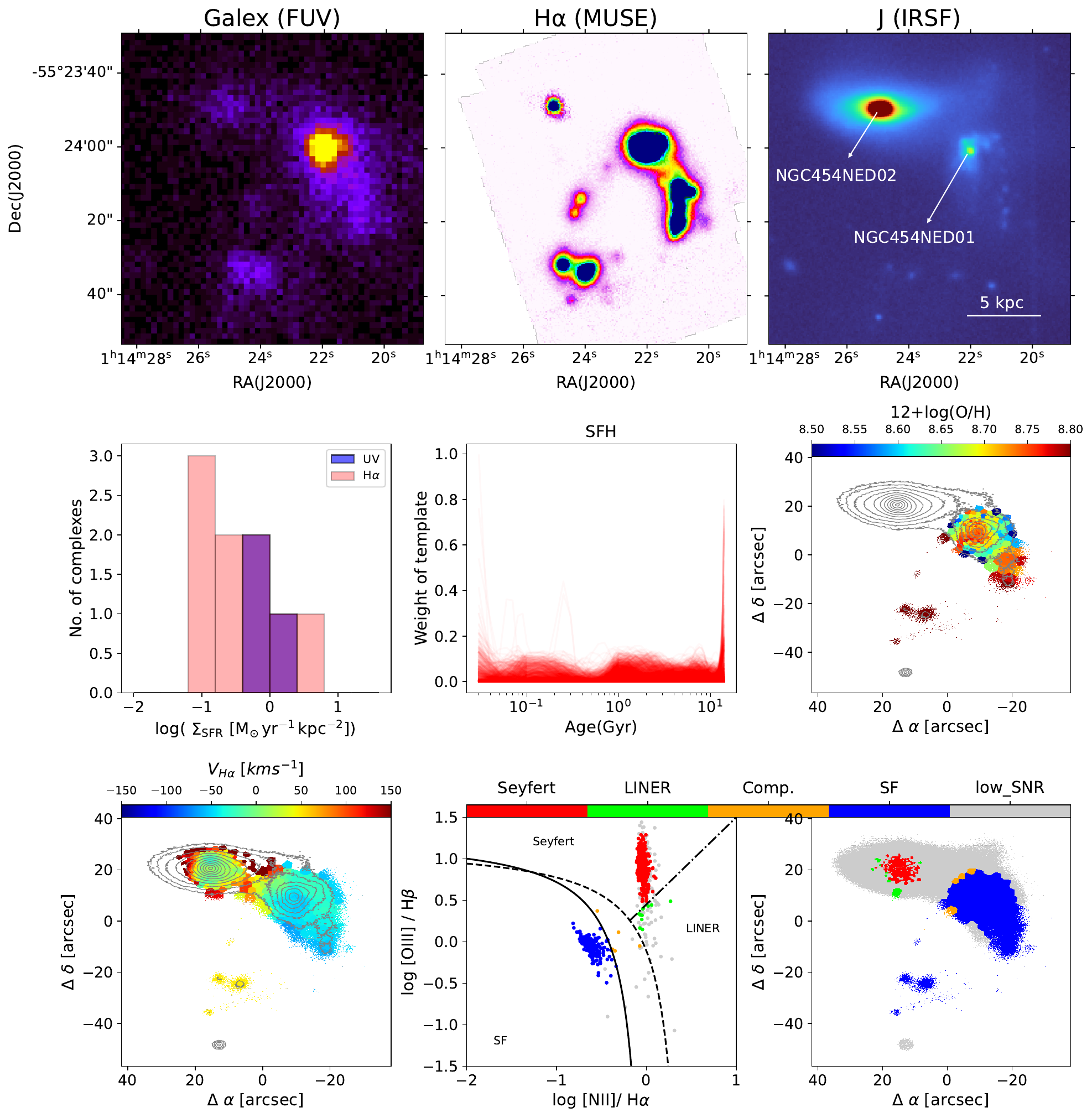}
    \caption{The top left, middle and right panels show the images of NGC454 in FUV, H$\alpha$ and J bands, respectively. The middle left panel shows  the histogram of $\Sigma_{SFR(FUV)}$ and $\Sigma_{SFR(H\alpha)}$. The middle and right panels show the SFH and metallicity. The bottom left panel shows the V$_{H_\alpha}$ velocity. The bottom right panel shows the BPT map of the galaxy.}
    \label{fig:NGC454}
\end{figure*}

\begin{figure*}
    \centering
    \includegraphics[width=0.8\textwidth]{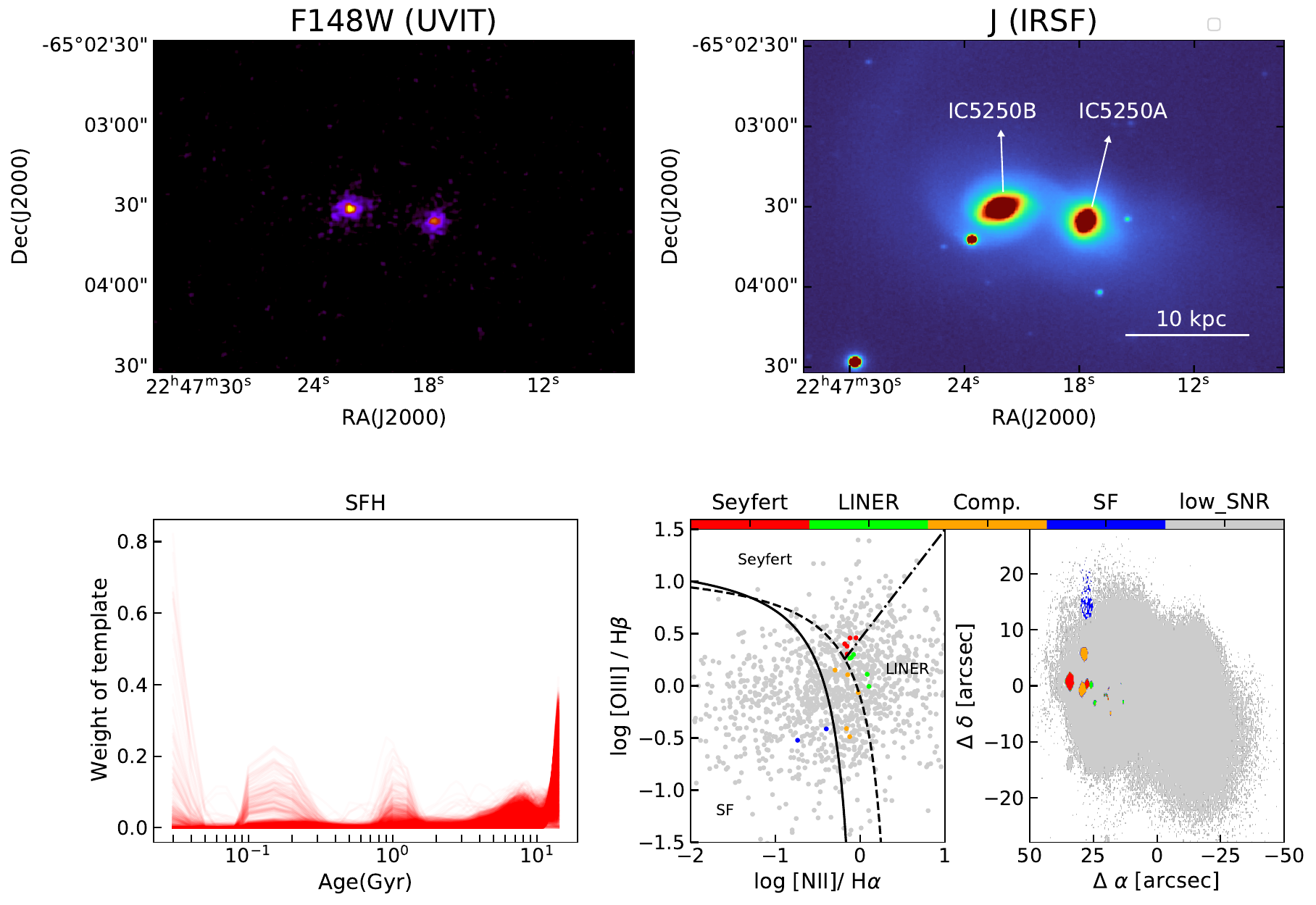}
    \includegraphics[width=0.8\textwidth]{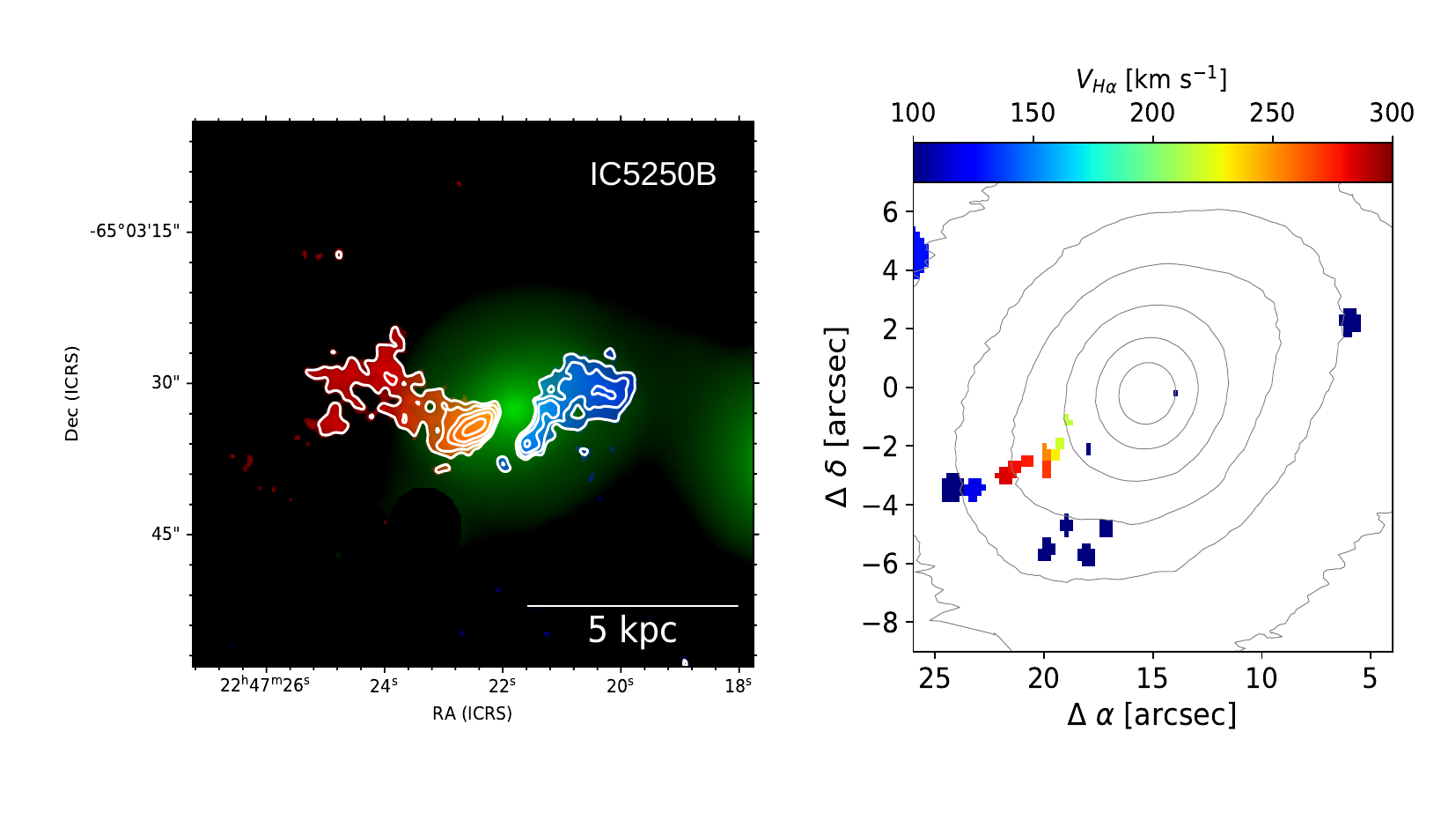}
    \caption{The top left and the middle panel shows images of IC5250 in FUV and J bands. The middle left and the middle panel show the metallicity and the BPT diagram. The bottom left panel shows the colour composite image of H$\alpha$. The blue and red colours indicate the blueshifted and redshifted sides of the outflow. The green background colour represents the white light image of IC5250. The white contours indicate the intensity of the blueshifted and redshifted outflows. The bottom right panel shows the H$\alpha$ velocity of the companion galaxy IC5250B lying east of IC250.} 
    \label{fig:IC5250}
\end{figure*}

\begin{figure*}
    \centering
    \includegraphics[width=0.75\textwidth]{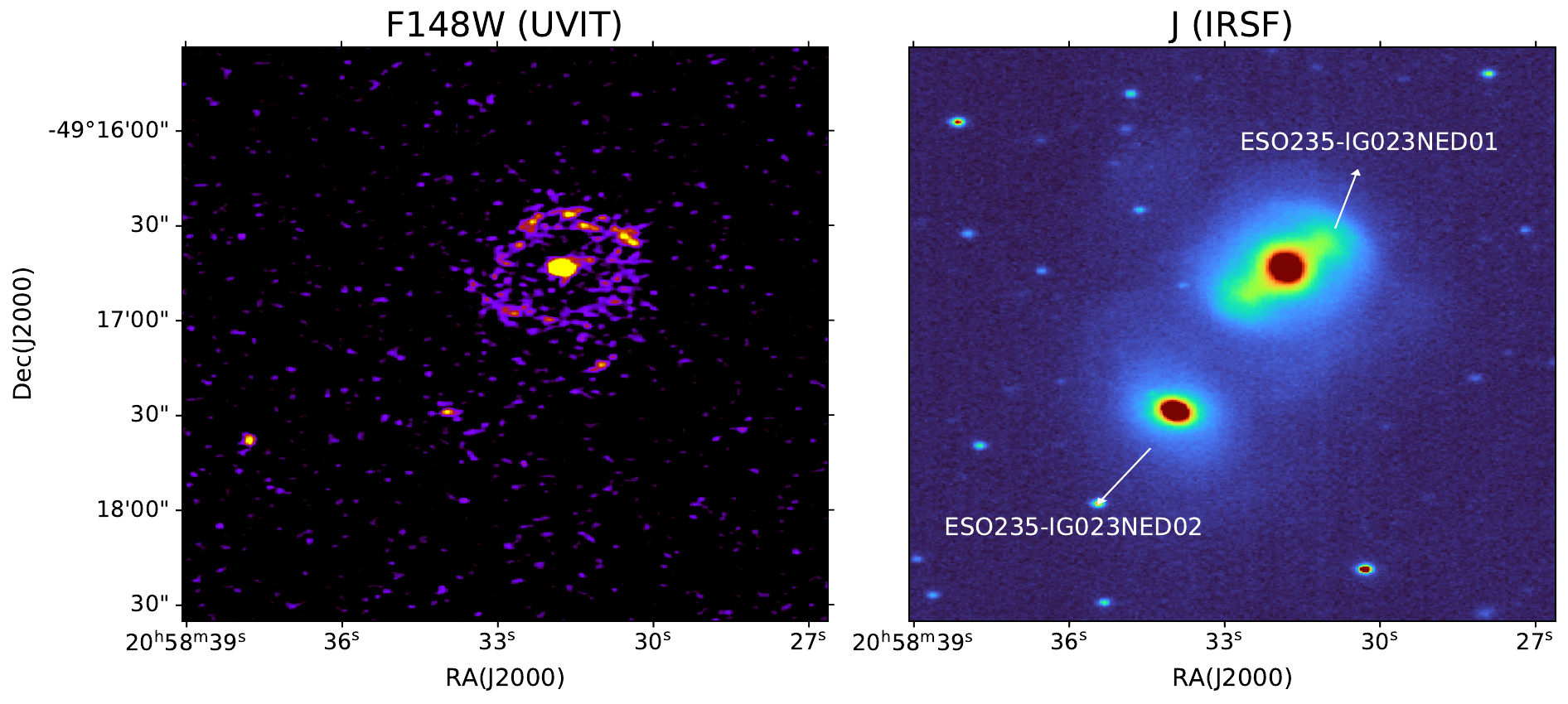}
    \caption{The left and right panels show the FUV and J band images of ESO235-IG023.}
    \label{fig:eso235}
\end{figure*}

\begin{figure*}
    \centering
    \includegraphics[width=0.7\textwidth]{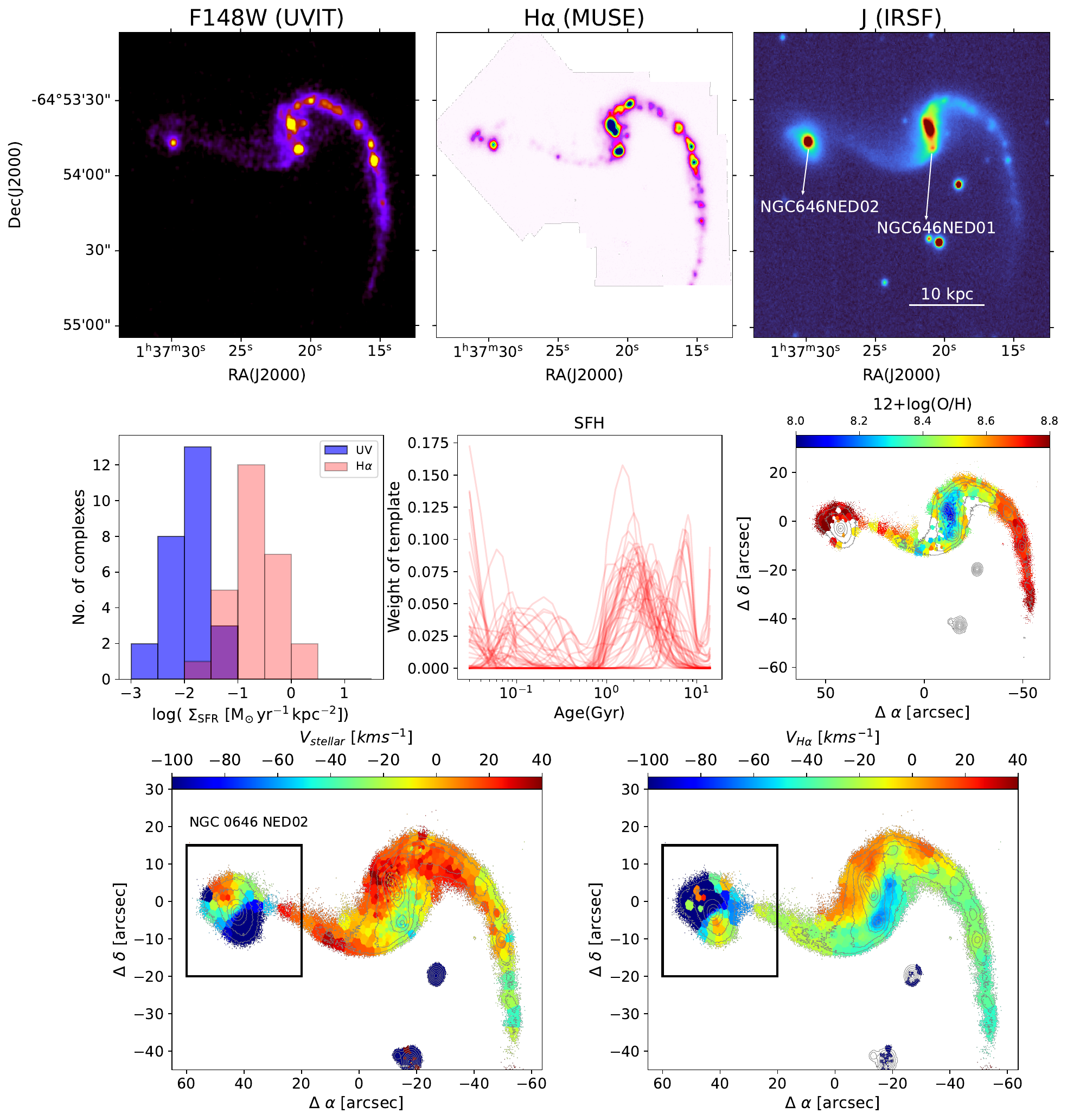}
    \hspace*{1.cm}\includegraphics[width=0.5\textwidth]{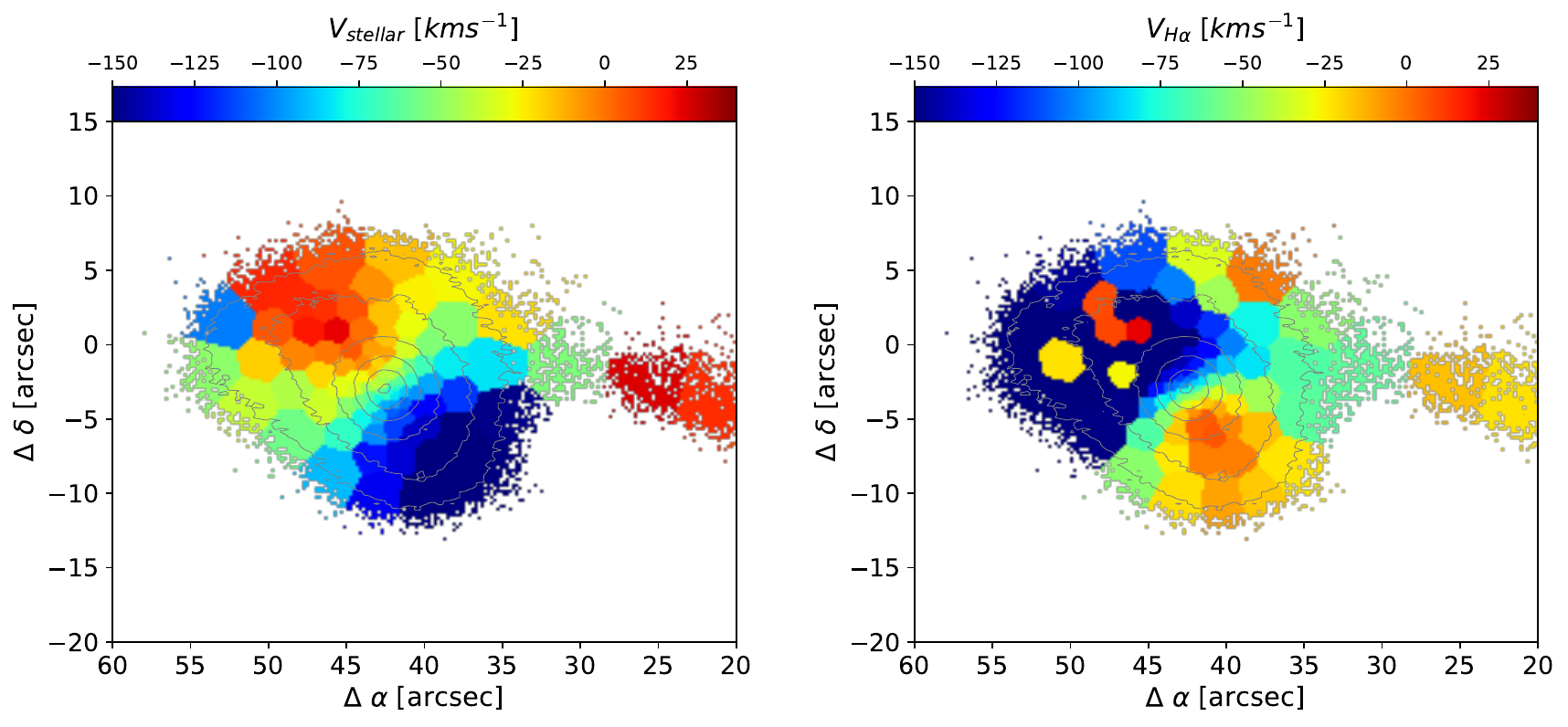}
    \hspace*{1.3cm}\includegraphics[width=0.95\textwidth]{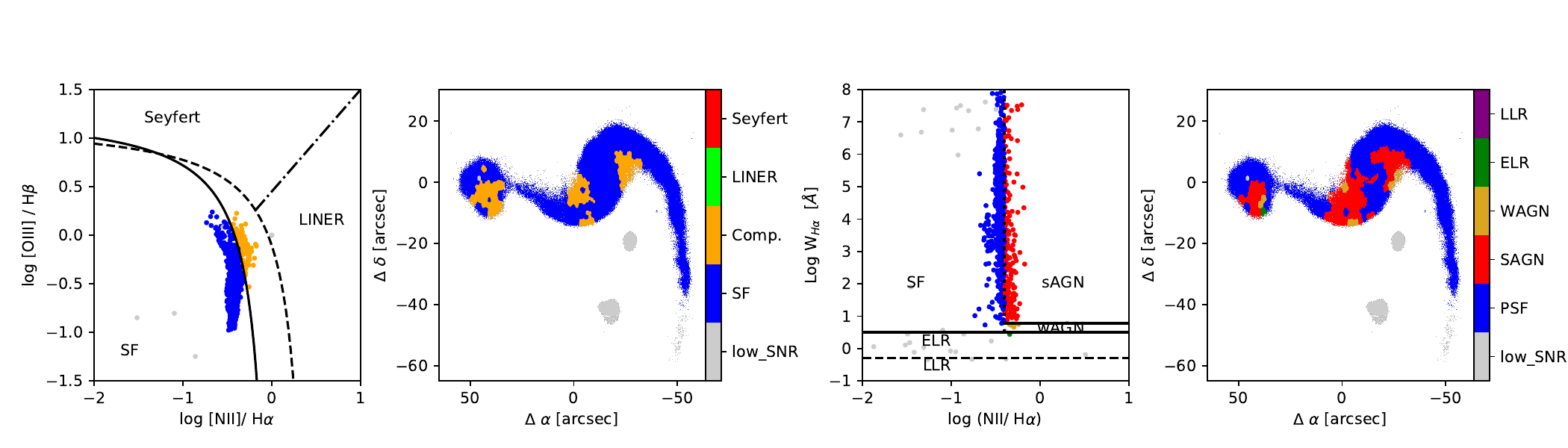}
    \caption{The top left, middle and right panels show the images of NGC646 in FUV, H$\alpha$ and J band, respectively. The 2$^{nd}$ row left panel shows the histogram of $\Sigma_{SFR(FUV)}$ and $\Sigma_{SFR(H\alpha)}$. The 2$^{nd}$ row middle and right panels show the SFH and metallicity. The 3$^{rd}$ row left and right panels show the V$_{stellar}$ and  V$_{H\alpha}$ of NGC646.
    The black rectangle shows NGC646NED02. The 4$^{th}$ row left and right panels show the V$_{stellar}$ and  V$_{H\alpha}$ of NGC646NED02. The bottom left and right panels show the BPT and WHAN diagram of the galaxy.}
    \label{fig:NGC646}
\end{figure*}

\begin{figure*}
    \centering
    \includegraphics[width=0.9\textwidth]{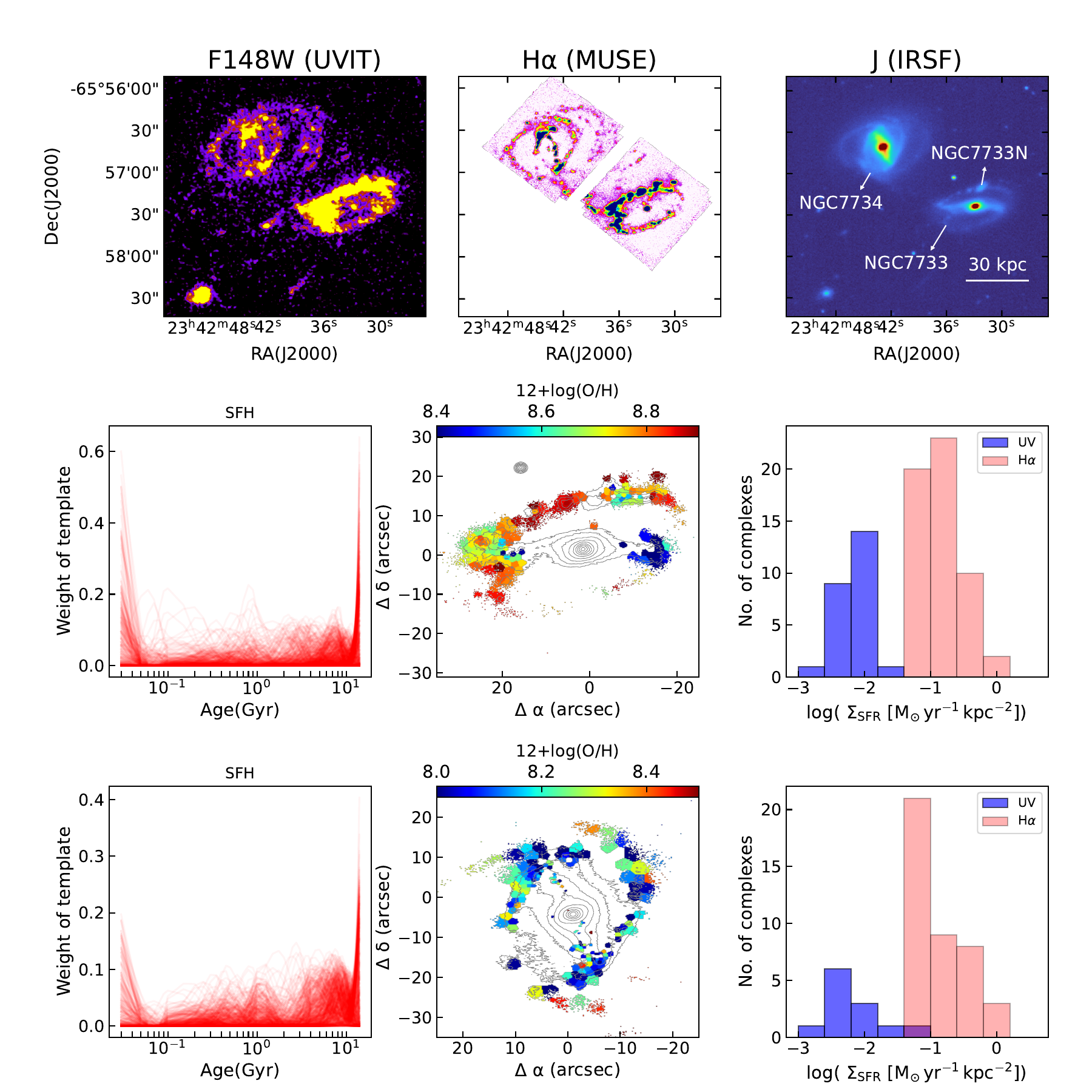}
    \includegraphics[width=0.25\textwidth]{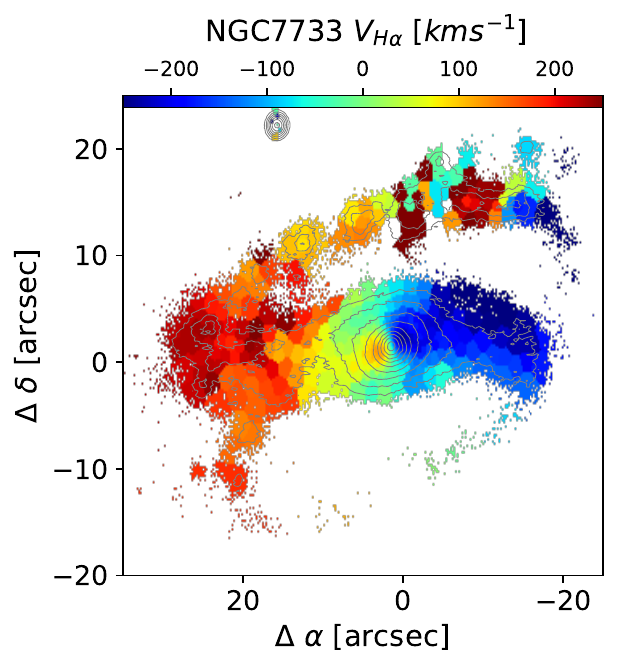}
    \includegraphics[width=0.25\textwidth]{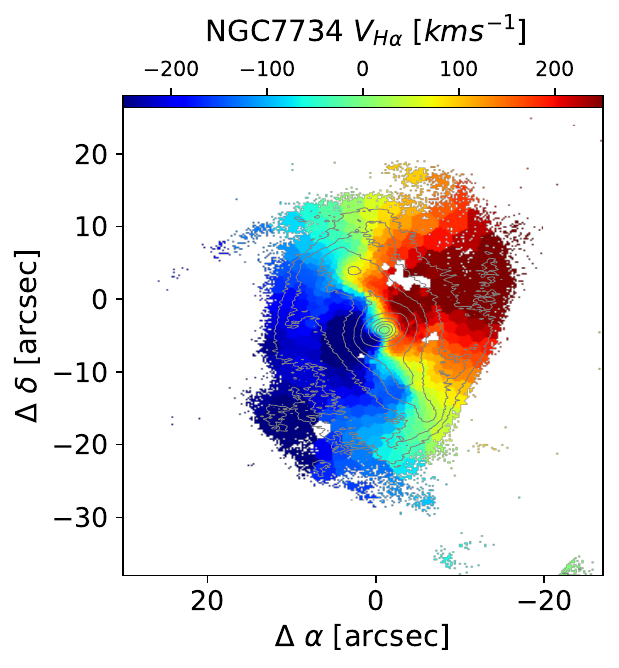}
     \includegraphics[width=0.35\textwidth]{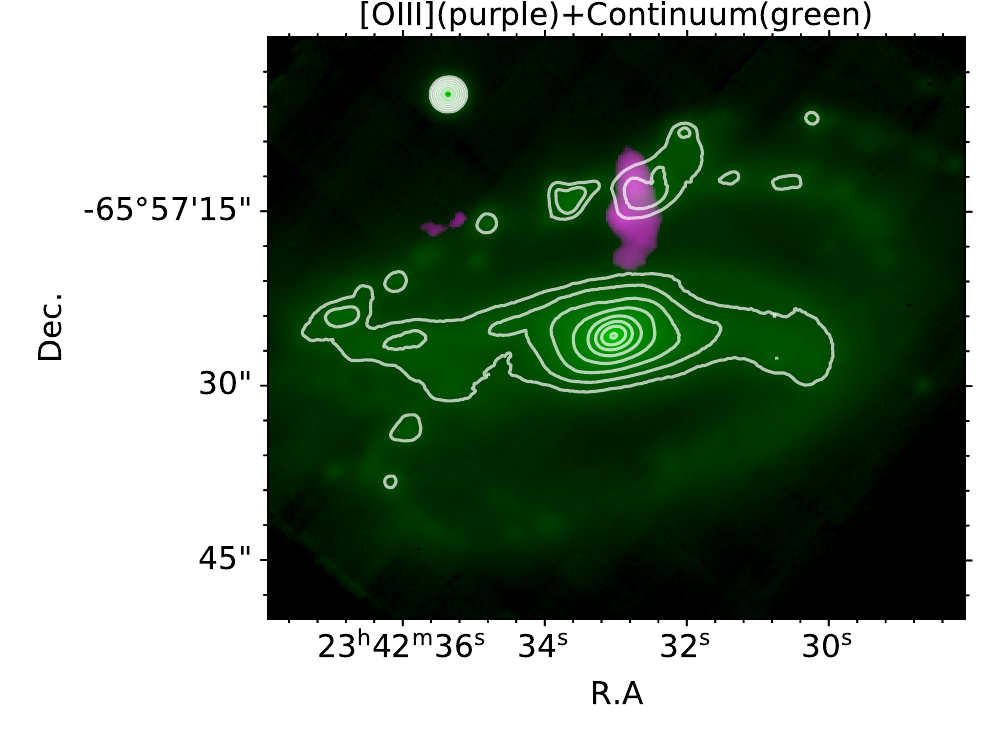}
    \caption{The top panel shows images of the NGC7733-34 system in FUV, H$\alpha$ and J band, respectively. The 2$^{nd}$ row left, middle and right panels show the SFH, metallicity and histogram of $\Sigma_{SFR}$ in FUV and H$\alpha$ of NGC7733. The 3$^{rd}$ row left, middle and right panels show the SFH, metallicity and histogram of $\Sigma_{SFR}$ in FUV and H$\alpha$ of NGC7734. The bottom left and middle panels show the V$_{H_\alpha}$ velocity of NGC7733 and NGC7734, respectively. The bottom right pane shows the colour-combined image of the white light image (green) and [\ion{O}{iii}](purple). The OIII emission shows an outflow from the central region of NGC7733N.}
    \label{fig:NGC7733}
\end{figure*}

\begin{figure*}
    \centering
    \includegraphics[width=0.8\textwidth]{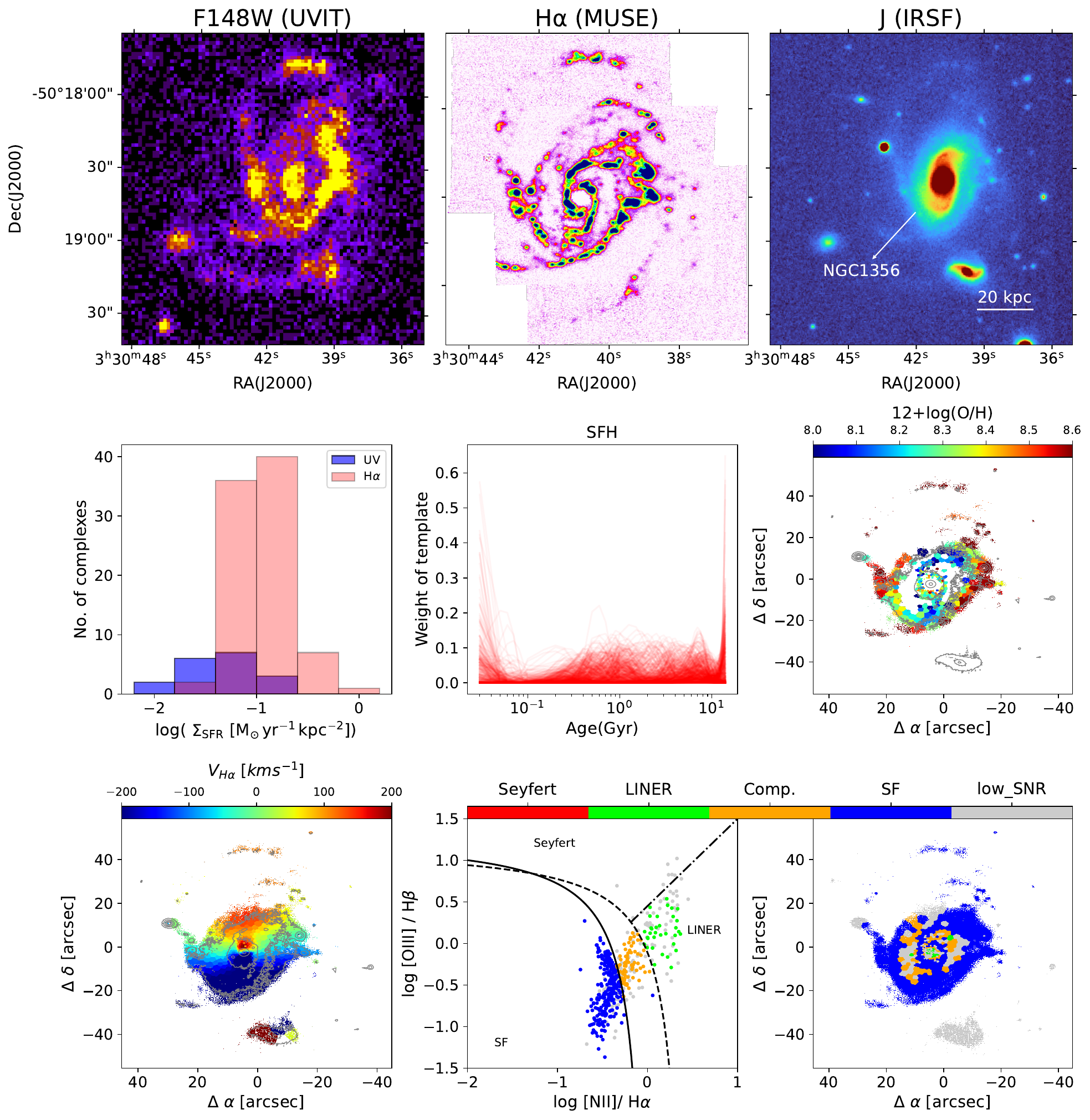}
    \caption{The top left, middle and right panels show the image of NGC1356 in FUV, H$\alpha$ and J band, respectively. The middle left panel shows  the histogram of $\Sigma_{SFR(FUV)}$ and $\Sigma_{SFR(H\alpha)}$. The middle and right panels show the SFH and metallicity. The bottom left panel shows the V$_{H_\alpha}$ velocity. The bottom right panel shows the BPT map of the galaxy.}
    \label{fig:NGC1356}
\end{figure*}

\begin{figure*}
    \centering
    \includegraphics[width=0.9\textwidth]{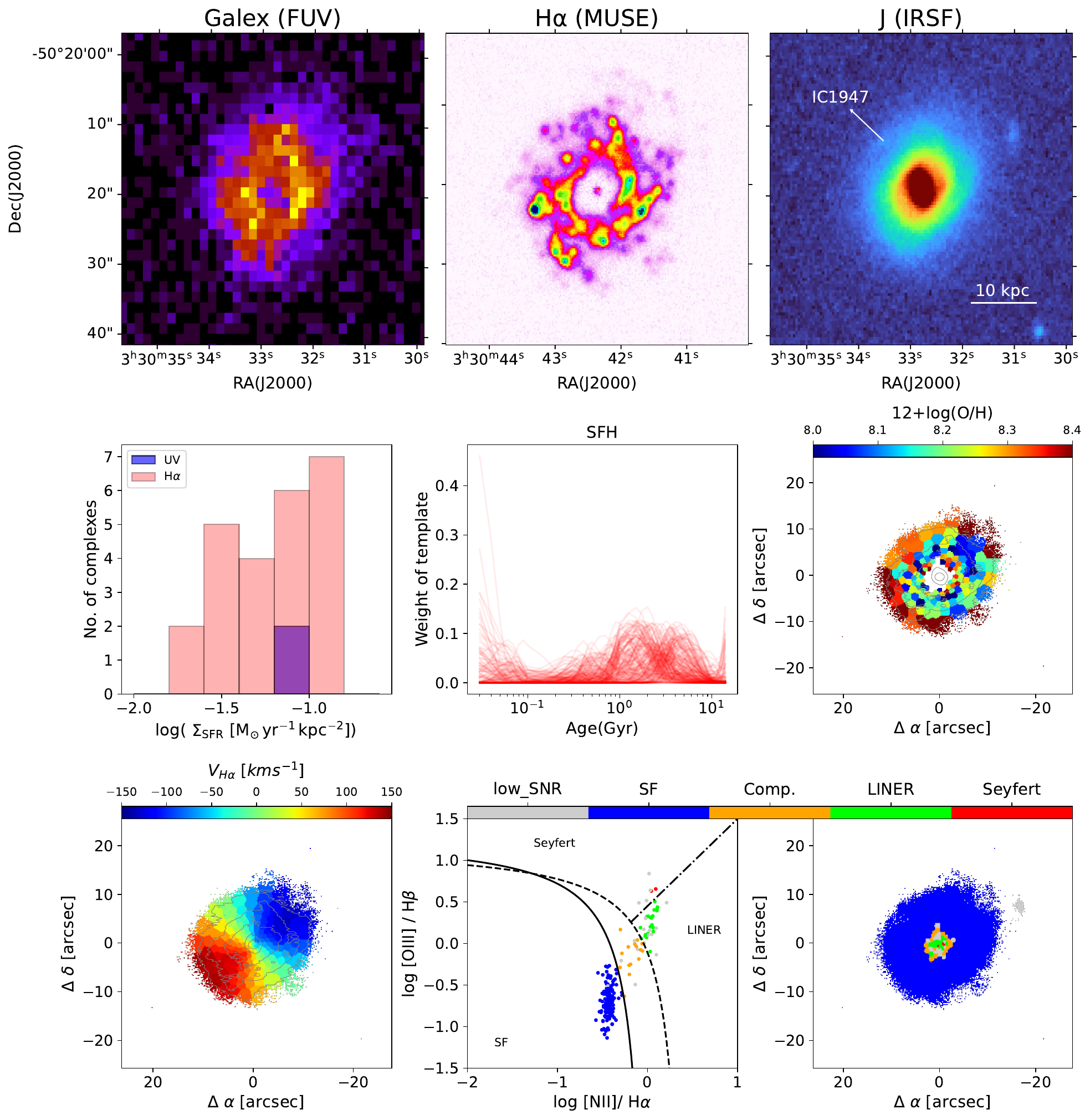}
     \includegraphics[width=0.7\textwidth]{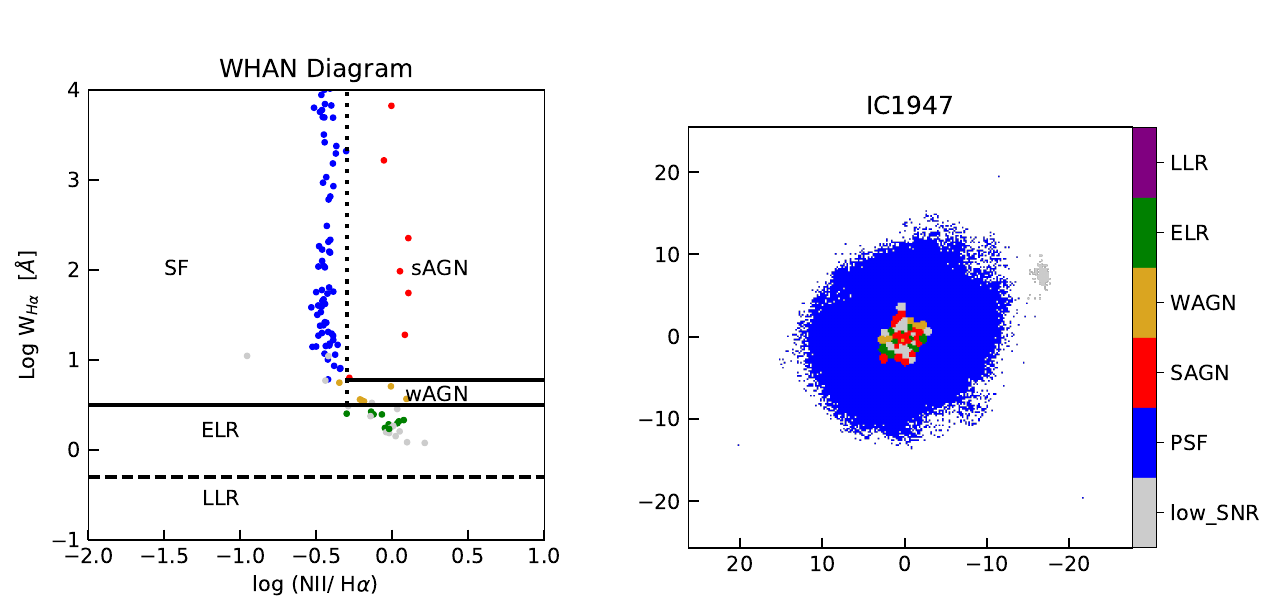}
    \caption{The top left, middle and right panels show the image of IC1947 in FUV, H$\alpha$ and J band, respectively. The 2$^{nd}$ row left panel shows the histogram of $\Sigma_{SFR(FUV)}$ and $\Sigma_{SFR(H\alpha)}$. The 2$^{nd}$ row middle and right panels show the SFH and metallicity. The 3$^{rd}$ row left and right panels show the V$_{H_\alpha}$ velocity and BPT diagram. The bottom right panel shows the BPT map of the galaxy. The bottom panel shows the WHAN diagram of IC1947.}
    \label{fig:IC1947}
\end{figure*}

\begin{figure}
    \centering
\includegraphics[width=0.47\textwidth]{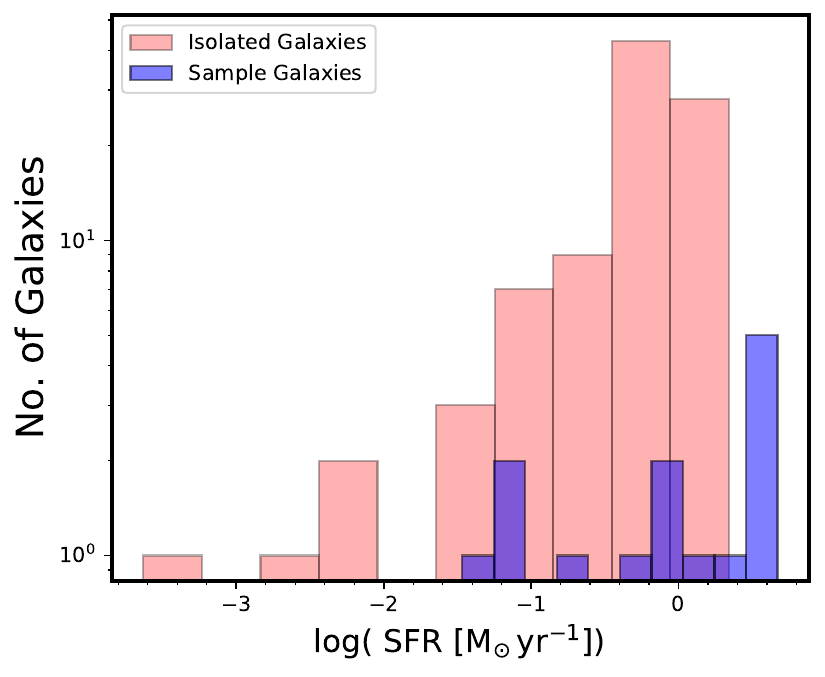}
    \caption{Histogram shows the comparison of sample galaxies with isolated galaxies \citep{Karachentsev2013AstBu..68..243K}.}
    \label{fig:comp_isolated}
\end{figure}

\begin{figure*}
    \centering
    \includegraphics[width=0.98\textwidth, height=0.8\textwidth]{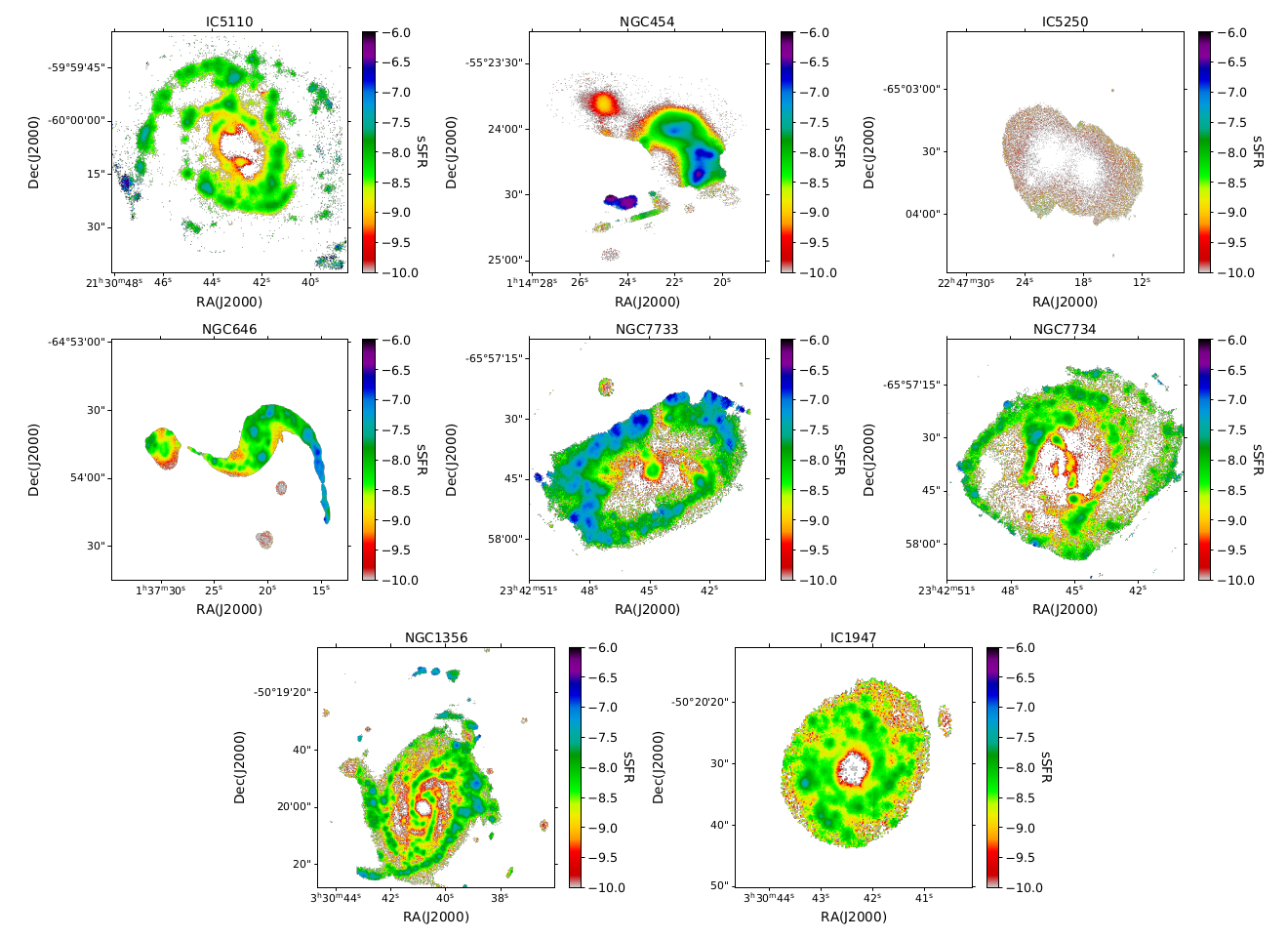}
    \caption{The sSFR maps of galaxies in the sample. The galaxies are star-forming, except IC5250, which does not show star formation and lies in the passive region.}
    \label{fig:ssfr}
\end{figure*}

\subsection{Specific star formation rate}
The sSFR explicitly shows how star formation contributes to galaxy growth and the relation between stellar mass and SFR. It is defined as SFR per unit stellar mass. Several observational studies show a correlation between SFR and stellar mass of star-forming galaxies \citep{Noeske2007ApJ...660L..47N, Elbaz2007A&A...468...33E,Elbaz2011A&A...533A.119E,Karim2011ApJ...730...61K}. \citet{Elbaz2007A&A...468...33E} showed the relation between stellar mass and SFR scales with cosmic time. The SFR of galaxies increases with redshift for a given stellar mass. The same evolution can also be seen in the sSFR of galaxies. The sSFR of galaxies also increases with redshift (up to z$\sim$2) for a given stellar mass. \citet{Bauer2005ApJ...621L..89B} found that sSFR decreases as the galaxy's stellar mass increases, which suggests that star formation is more active and contributes more to the growth of low-mass galaxies while the more massive galaxies formed their stars in the past. We created the SFR maps of each galaxy using H$\alpha$ maps. We estimated the stellar mass of the galaxies in our sample using the method given in \citet{Bell2003ApJS..149..289B}. 
\begin{equation}
    log10(M/L_{i})= a_{i}+b_{i}\times(r-i)
\end{equation}
The values of a and b are 0.006 and 1.114 respectively. We used pyMuse to create r, and i band images of the individual galaxy from MUSE cubes. The sSFR of the galaxies is shown in Fig.~\ref{fig:ssfr}.


\section{Results and discussion}

\subsection{Effect of interaction on star formation and kinematics}

Galaxy interaction can be an effective mechanism to regulate star formation. All the galaxies in our sample are gas-rich and show recent and massive star formation activity in their disk except IC5250. The ongoing star formation is also evident from FUV and H$\alpha$. The galaxies in our sample show complex spiral structures, rings, tidal tails, and bridges hosting star-forming regions. 

\subsubsection{ESO341-IG004}
ESO341-IG004 is in an advanced stage of merging, with plumes and tidal tails extending towards the nucleus's east and west. These tidal features could be due to the gas stripping of another galaxy that has merged with ESO341-IG004. The galaxy shows a loop on the southwestern side in the NIR map. The old stellar bulge is also prominent in the NIR image. The galaxy also shows plumes having bright knots of recent star formation on the eastern side in the FUV map (Fig.~\ref{fig:eso341}). These massive SFCs could have been formed during the tidal stripping of a minor galaxy. The galaxy has a mean FUV $\Sigma_{SFR}$ of 0.209 (M$_\odot$ yr$^{-1}$ kpc$^{-2}$). MUSE data is unavailable for this system. Thus, we could not look into other properties.

\subsubsection{IC5110}
IC5110 represents an advanced merger system, where the bulges of the two galaxies are in the final stage of merging. The J-band images show two bulges in the centre. The third bright point in the centre is a foreground star. IC5110 shows star formation along the arms, evident from the FUV and H$\alpha$ images (Fig.~\ref{fig:IC5110}). The galaxy has a mean FUV and H$\alpha$ $\Sigma_{SFR}$ of 0.007 and 0.061 (M$_\odot$ yr$^{-1}$ kpc$^{-2}$) respectively. The SFH map of (Fig.~\ref{fig:IC5110}) shows recent star formation activity. The galaxy did not form many stars in the previous 1Gyr, and the merger activity could have triggered recent star formation activity. Interactions can exert torques on the gas and induce bar formation or enhance the bar strength in the main galaxy. The gas outside the corotation radius can gain angular momentum and be driven outwards to form rings at outer resonances. Since IC5110 shows an inner ring and outer rings in FUV and H$\alpha$, the gas may have been subjected to positive torques due to the interaction, which has led to resonant rings formation in IC5110.

\subsubsection{NGC454}
NGC454 is a strongly interacting system and is composed of two components. The western component NGC454W shows a huge amount of hot and massive star formation seen as bright knots in FUV and H$\alpha$. However, the eastern companion, NGC454E, is an early type galaxy and faint in H$\alpha$ and FUV (Fig.~\ref{fig:NGC454}). NGC454NED02 (NGC454E) contains an old stellar population and dust lanes and is bright in NIR. \citet{Johansson1988A&A...191...29J} presented NGC454NED02 as a red elliptical galaxy. Later on, \citep{Stiavelli1998ApJ...492L.135S} showed that NGC454NED02 is an S0 galaxy. The luminosity profile can be fitted with two components, an exponential profile \citep{Freeman1970ApJ...160..811F} for a disk in addition to an r$^{1/4}$ profile \citep{deVaucouleurs1948AnAp...11..247D}. \citet{Marchese2012MNRAS.421.1803M} showed that the nucleus in NGC454NED02 is a changing look AGN, which has been observed in both Compton-thin and reflection-dominated states. The western component NGC454W is highly distorted due to the interaction. It was classified as the debris of an irregular galaxy by \citet{Johansson1988A&A...191...29J}  while \citet{Stiavelli1998ApJ...492L.135S} using high-resolution \textit{Hubble Space Telescope} (\textit{HST}) images showed that it is the debris of a disk galaxy. Using smooth particle hydrodynamic simulations, \citet{Plana2017MNRAS.472.3074P} suggested that the system is 12.4 Gyr old and is the result of interaction between two halos having equal mass and perpendicular spin. NGC454 shows previous episodes of star formation. The mean FUV and H$\alpha$ $\Sigma_{SFR}$ of NGC454NED01 is 0.915 and 0.427 (M$_\odot$ yr$^{-1}$ kpc$^{-2}$) respectively (Fig.~\ref{fig:NGC454}).) There is higher metallicity in the disk regions, with a metallicity range (12 + log(O/H)) of 8.5--8.75 (Fig.~\ref{fig:NGC454}).

\subsubsection{IC5250}
IC5250 is an advanced merger system composed of two interacting galaxies, IC5250A and IC5250B. IC5250A and IC5250B are S and S0 types respectively \citep{Donzelli2000AJ....120..189D}. The 
 FUV image shows bright emission in the central region, associated with the hot, low mass, evolved population of stars. The extent of the system is much larger in infrared. The system also shows tidal tails, which are bright in infrared (Fig.~\ref{fig:IC5250}). IC5250 is gas-poor, and there is no star formation. Thus, we could not determine the gas phase metallicity of this galaxy.

\subsubsection{ESO235-IG023}
ESO235-IG023 system is composed of two galaxies, ESO235-IG023NED01 and ESO235-IG023NED02. These galaxies share the same envelope. The southern galaxy ESO235-IG023NED02 shows a prominent bulge in the infrared map, which is faint in the FUV map. The minor galaxy ESO235-IG023NED02 is faint in the FUV map (Fig.~\ref{fig:eso235}). ESO235-IG023 shows a bar in the NIR and a ring near the corotation of the bar in UVIT. These rings form due to the slow accumulation of gas at the resonances due to the gravitational torques applied by the bar. The UVIT maps reveal several SFCs in the ring. However, the FUV data does not have enough S/N, and thus we could not detect SFCs above the 5$\sigma$ threshold. We have calculated the $\Sigma_{SFR}$ of ESO235NED01 in FUV by taking an elliptical annulus around the ring. MUSE data is unavailable for this galaxy, so we could not look into the metallicity, SFH and velocity profiles of the interacting system.

\subsubsection{NGC646}
NGC646 is a double system composed of two interacting galaxies. The western component, NGC646NED01, shows two arms. The western arm of NGC646NED01 shows bright star-forming knots in FUV and H$\alpha$. These star-forming regions are also bright in the NIR image (Fig.~\ref{fig:NGC646}). The eastern arm of NGC646NED01 is distorted and appears smooth and diffuse. NGC646NED01 also has a short bar \citep{deVaucouleurs1991rc3..book.....D}. A satellite galaxy, NGC646NED02, is attached to the eastern arm of NGC646NED01. The system shows a recent episode of star formation. The mean FUV and H$\alpha$ $\Sigma_{SFR}$ of NGC646NED01 is 0.015 and 0.451 (M$_\odot$ yr$^{-1}$ kpc$^{-2}$)respectively. The mean FUV and H$\alpha$ $\Sigma_{SFR}$ of NGC646NED02 is 0.022 and 0.260 respectively (Fig.~\ref{fig:NGC646}). The interaction can lead to strong morphological distortions \citep{Hota2007MNRAS.380.1009H}.  Studies of CALIFA merging galaxies show that ionised gas kinematics are more likely to react to interactions. The stellar and ionized gas velocity maps of NGC646NED02 show counter-rotating disks. Such counter rotation has been linked with external gas accretion \citep{Corsini2014ASPC..486...51C}. In these systems, ionized, molecular and atomic gas rotate with similar amplitude to the stellar body but in the opposite direction. The counter-rotation is also evident from the velocity and metallicity maps. The metallicity of the eastern arm of NGC646NED01 differs from that of NGC646NED02 (Fig. \ref{fig:NGC646}). The H$\alpha$ velocity of the eastern arm of NGC646NED01 differs from its stellar velocity, suggesting that gas is moving towards NGC646NED02 from NGC646NED01. This shows that the NGC646NED02 is accreting gas from the eastern arm of NG646NED01.

\subsubsection{NGC7733-34}
NGC7733-34 is a pair of interacting galaxies that show star formation knots in their inner disks, rings and tidal arms. Both NGC7733 and NGC7734 are spiral barred galaxies \citep{1991deVaucouleurs}. NGC7733 and NGC7734 are separated by a distance of 48.3 kpc. NGC7733 is also interacting with a minor galaxy, NGC7733N \citep{Yadav2021A&A...651L...9Y}. NGC7733 and NGC7734 show a recent burst of star formation. NGC7733N appears to be an early-type galaxy interacting with NGC7733 and NGC7734. We did not find any SFCs from NGC7733N. The mean FUV and H$\alpha$ $\Sigma_{SFR}$ of NGC7733 is 0.008 and 0.194 (M$_\odot$ yr$^{-1}$ kpc$^{-2}$)respectively. The mean FUV and H$\alpha$ $\Sigma_{SFR}$ of NGC7734 is 0.014 and 0.200 (M$_\odot$ yr$^{-1}$ kpc$^{-2}$)respectively (Fig.~\ref{fig:NGC7733}).  NGC7733 and NGC7734 are barred galaxies and show rings. Both NGC7733 and NGC7734 also show recent massive SFCs along the rings (Fig.~\ref{fig:NGC7733})

\subsubsection{NGC1356}
NGC1356 is an intermediate barred galaxy with prominent star formation along its spiral arms. The arms of the galaxy appear distorted due to the tidal interaction (Fig.~\ref{fig:NGC1356}). It is interacting with the galaxy IC 1947, which is present in the southwestern region. IC 1947 also shows ongoing star formation in the disk (Fig.~\ref{fig:IC1947}). The torques due to the bar can drive the gas radially, inward or outward from corotation, and lead to nuclear, inner and outer rings. Nuclear rings are present near the nuclear region of the galaxy at the inner Lindblad resonance, while the inner rings encircle the bar and are elongated along the bar axis \citep{Buta1999Ap&SS.269...79B}. NGC1356 also has a nuclear ring, and IC1947 has an inner ring hosting several SFCs. The mean FUV and H$\alpha$ $\Sigma_{SFR}$ of NGC1356 is 0.058 and 0.131 (M$_\odot$ yr$^{-1}$ kpc$^{-2}$) respectively. The mean FUV and H$\alpha$ $\Sigma_{SFR}$ of IC1947 is 0.082 and 0.073 (M$_\odot$ yr$^{-1}$ kpc$^{-2}$) respectively (Fig.~\ref{fig:NGC1356} and Fig.~\ref{fig:IC1947}). It is also clear from the MUSE images that NGC1356 and IC1947 currently have active episodes of star formation. NGC1356 and IC1947 show higher metallicity along the arms.

\subsection{The distribution of sSFR and SFCs over the sample galaxies}

The galaxies in the sample show higher sSFR in the spiral arms than inter-arm regions Fig.~\ref{fig:ssfr}, which is expected, as spiral arms are the sites of higher gas than stellar concentration. NGC454 shows the highest sSFR in the tidal tails among the sample galaxies. NGC7733 shows higher sSFR in the northeast region, which is closer to NGC7734. NGC646NED01 also shows enhancement in sSFR along the eastern tidal tail. IC5250A and IC5250B are passive galaxies which show low sSFR. The sSFR values of the galaxies in the sample except IC5250 lie in the star-forming region, Fig. \ref{fig:ssfr}. This suggests that the interaction of gas-rich galaxies triggers massive star formation, which helps in building the stellar mass of galaxies. The ongoing star formation in the sample of interacting galaxies suggests that the tidal forces between galaxies during galaxy interactions may compress the gas and lead to star formation.

Fig.~\ref{fig:comp_isolated} shows the comparison of the galaxies in this study with isolated galaxies from \citet{Karachentsev2013AstBu..68..243K}. We selected only the galaxies from the isolated sample which have log($M_{\star}$) >10 to match the stellar mass with the sample galaxies. The mean log(SFR) for the sample galaxy is $-$0.077 M$_\odot$ yr$^{-1}$ and the mean log(SFR) for the isolated sample is $-$0.374 M$_\odot$ yr$^{-1}$. We have also performed the KS test (statistic=0.43, pvalue=0.015), suggesting that both interacting and isolated galaxies follow different distributions. This suggests that galaxies in the sample show enhancement in $\Sigma_{SFR}$ with respect to isolated galaxies. We have also looked at the $\Sigma_{SFR}$ of galaxies as a function of distance.

\begin{figure}
    \centering
\includegraphics[width=0.45\textwidth]{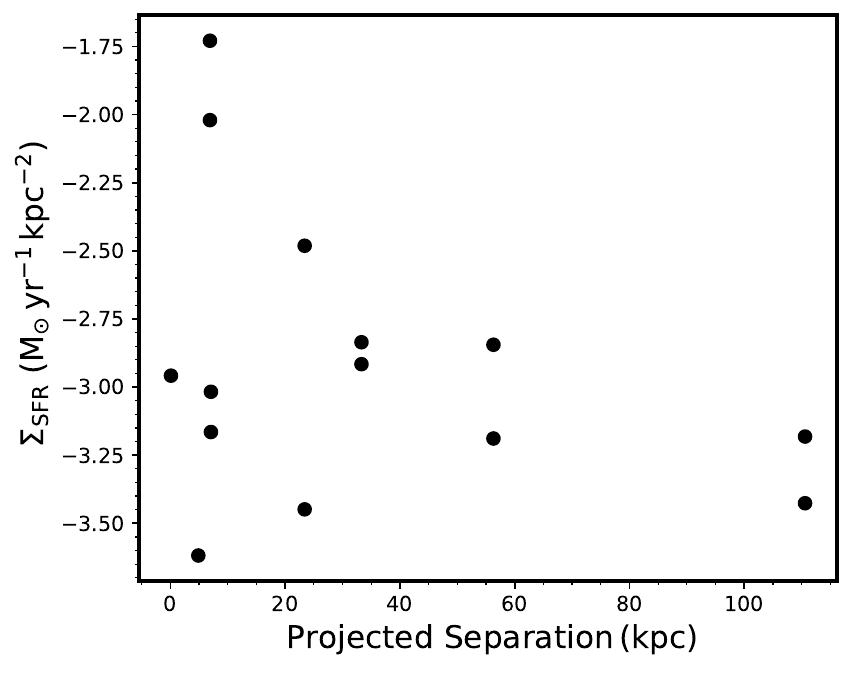}
    \caption{$\Sigma_{SFR(FUV)}$ of the sample galaxies with respect to the projected separation between the interacting galaxies.} 
    \label{fig:sfr_distance}
\end{figure}

Fig. \ref{fig:sfr_distance} shows the $\Sigma_{SFR}$ in the sample galaxies as a function of projected separation between them. We did not find any correlation of $\Sigma_{SFR}$ with the projected separation of interacting pairs, although this could be due to the small sample size. The enhancement in the $\Sigma_{SFR}$ of a galaxy depend on various factors such as the distance between galaxies, mass ratio, gas content and the types of galaxies involved in the interaction. Change in $\Sigma_{SFR}$ is investigated in relation to the projected distance, as it is considered an indication of the merging phase of a pair of galaxies. When galaxies are closer, it is more likely that they are undergoing a close encounter. Conversely, when galaxies are farther apart, they may either approach each other or move away after their initial closest approach. However, confirming these scenarios is challenging because the projected distance merely offers a snapshot of the interaction and does not provide direct information about the time frame involved.

Fig. \ref{fig:NGC7734_color_sfr} shows the extracted SFCs in NGC7734, which are coloured based on their $\Sigma_{SFR(H\alpha)}$. The SFCs along and at the edges of the bar show higher $\Sigma_{SFR}$ compared to disk SFCs. This suggests that the bar may help in funnelling the gas towards the centre. The number of detected SFCs is larger in H$\alpha$ than in FUV. However, FUV can trace an older stellar population (100Myr) compared to H$\alpha$ (10Myr), although the FUV observations are not as deep as the H$\alpha$ images. The dust extinction is also high in FUV. Thus we may have missed some SFCs in FUV. Dust extinction may vary from galaxy to galaxy and within a galaxy. This can also affect the number of FUV SFCs detected. 
Observation studies using MaNGA data show the enhancement in star formation activity after the first pericenter passage \citep{Pan2019ApJ...881..119P}. The galaxies in the sample show rising star formation activity in star formation history maps. The recent star formation activity is also evident from FUV and H$\alpha$ maps. 
\citet{Gonz2017A&A...607A.128G} have looked at the spatially resolved star formation history of different Hubble-type galaxies using CALIFA data. They suggested that early-type galaxies (E and S0) grow their envelope and their inner regions through gas accretion or mergers.
The resolved sSFR map of our sample galaxies shows higher sSFR along the arms and tails, except for IC5250. 
Enhancement in sSFR of interacting galaxies has also been shown in multiple studies \citep{Bell2006ApJ...652..270B, Jogee2009ApJ...697.1971J, Robaina2009ApJ...704..324R, Ballesteros2015A&A...579A..45B} which suggests that mergers do trigger the star formation in galaxies.

The galaxies in our sample have prominent bulges which contain old and evolved populations of stars. The metallicity maps shown in Fig. \ref{fig:IC5110} to Fig. \ref{fig:IC1947} are consistent with photoionization in the BPT diagram. \citet{Ballesteros2015A&A...579A..45B} have used data from the CALIFA survey and found a moderate enhancement in the metallicity of the central regions of the interacting galaxies. Their result suggests moderately suppressed star formation in the outer regions.
The metallicity maps of our sample galaxies suggest the enhancement in the metallicity along the arms, rings and tails, which are the sites of recent and massive star-forming regions. This suggests that the stellar feedback from the massive stars can contribute to the enrichment. We do not see an enhancement in the metallicity of central regions. This has also been shown in multiple observational and numerical studies \citep{Ellison2008AJ....135.1877E, Yuan2012A&A...548A.117Y, Hopkins2013MNRAS.430.1901H, Moreno2015MNRAS.448.1107M}.

This suggests that the interaction of gas-rich galaxies can trigger star formation along the spiral arms, rings and tidal tails. The galaxies in the sample show rising star formation at their current phase in the SFH plot. The galaxies also show multiple bursts of star formation in the past. The ongoing star formation is also evident from FUV and H$\alpha$ maps. 


 \subsection{The nature of the nuclei}
The tidal torques during galaxy interaction can lead to gas inflows, which can also feed a central black hole and trigger AGN activity \citep{Sanders1988ApJ...325...74S}. We investigated the nature of the nuclei in each galaxy. We used the BPT and WHAN diagrams to check the source of ionization in the centre of the galaxies. 
The emission from the central region of NGC454 falls in the Seyfert region, which has also been suggested by \citet{Marchese2012MNRAS.421.1803M}. IC5250 also shows AGN activity from the central region. NGC7733 and NGC7733N host Seyfert nuclei, while NGC7734 hosts LINER nuclei. NGC7733 shows an extended narrow line region out to approximately 18 kpc \citep{Yadav2021A&A...651L...9Y}.
MUSE data for ESO341-IG004 and ESO235-IG023 is not available. Hence, we could not check the nature of nuclei in the centres of these galaxies. 
The origin of AGN activity in interacting/merging galaxies is still debated. AGN activity is a stochastic process which can vary over a timescale shorter than other galaxy-wide processes such as star formation or morphology. Multiple studies using observations and simulations have tried to understand the connection between interaction/mergers with the AGN activity. \citet{Gao2020A&A...637A..94G} have shown that AGN is more likely to be found in mergers, and the ratio of AGN fraction in mergers relative to non-mergers is 1.81. Studies using MaNGA data show that tidal interactions can induce around 30 $\%$ more AGN than stochastically fueled AGN at projected separations of 10kpc \citep{Fu2018ApJ...856...93F, Steffen2023ApJ...942..107S}. Merger simulations also show enhanced AGN activity \citep{Mihos1996ApJ...464..641M, Rupke2010ApJ...710L.156R, Hopkins2013MNRAS.430.1901H, Moreno2019MNRAS.485.1320M}. Our sample consists of 8 interacting systems comprising of 15 galaxies. We found that six galaxies host AGN activity, and four do not host AGN in their central region. The nature of nuclear activity is unclear for two galaxies, and three galaxies do not have MUSE data. These numbers suggest that interaction/merger may help in fuelling the central supermassive black hole in galaxies and hence triggering AGN activity.

\subsubsection{The Outflow in IC5250B}
The ionized gas shows a different velocity than the stellar velocity near the nuclear region of IC5250B. In Fig. \ref{fig:IC5250}, the blue and red colours indicate the blueshifted and redshifted sides of the H$\alpha$ outflow. The background green emission represents the white light image of IC5250. The redshifted outflow has an extent of $\sim$5 kpc and an H$\alpha$ velocity up to $\sim$ 300 km s$^{-1}$. Both the galaxies in the system do not show significant gas and star formation activity. The velocity of outflowing gas is relatively low and perhaps not high enough to disperse gas from the galaxy. It is possible that the gas had already disappeared quite early, and it is a dry interaction between two lenticulars. We estimated the bulge to total mass ratio (B/T) of IC5250B by performing the 2D decomposition of the galaxy using Galfit \citep{Peng2002AJ....124..266P} on K band data. The B/T ratio of IC5250B is 0.95, which suggests that the bulge is massive in the galaxy. The remaining tracers of gas could currently fuel the outflow in IC520B.

\subsubsection{Nature of emission in NGC646}
 NGC646NED01 shows composite emission along the inside curved region of the bar end, and NGC646NED02 shows composite emission from the central region in the BPT diagram (Fig. \ref{fig:NGC646} bottom left panel). WHAN diagram shows that the composite emission along the curved region of the bar end from both galaxies is associated with the AGN activity (Fig. \ref{fig:NGC646} bottom right panel). The composite emission in the BPT diagram and strong AGN emission in the WHAN diagram is also arising in the disk, away from the nucleus. The strong shocks during the interaction process can also lead to high ionization, which may appear as a strong AGN in the WHAN diagram. (Fig. \ref{fig:NGC646} bottom right panel).
  We do not clearly understand the nature of nuclear activity in this system.

\citet{Monreal2010A&A...517A..28M} have used diagnostic diagrams and suggested that shocks cause ionization in extra-nuclear regions of interacting and merging galaxies. 
 \citet{Allen2008ApJS..178...20A} have also suggested that shocks can move the emission ratio to the LINERs part of the BPT diagram, and shock models better fit the LINER sequence. This suggests that tidal forces play an important role in the origin of ionizing shocks during the interaction of galaxies. 

 \subsubsection{Outflow in NGC7733N}
Galaxy interaction can have a dramatic effect on the host galaxy. The tidal forces can lead the gas flow towards the central region, igniting accretion onto the central supermassive black hole. NGC7733 and NGC7733N show AGN activity from their central region \citep{Yadav2021A&A...651L...9Y}.
 Fig. \ref{fig:NGC7733} bottom right panel shows the colour composite image of white light image (green) and [\ion{O}{iii}] image (purple). The [\ion{O}{iii}] emission at the location of NGC7733N suggests the presence of an outflow. The total extent of the [\ion{O}{iii}] outflow is around 8kpc, which is comparable to the size of the galaxy.   

 \subsubsection{AGN in IC1947}
Some of the galaxies in our sample show LINER emission from their central region in the BPT diagram. This is the case for IC5110, NGC1356 and IC1947. The LINER emission can also arise due to the hot low mass evolved population of stars. These evolved populations of stars can ionize the gas and produce a line ratio similar to that of LINER. We find that the emission from the central region of only IC1947 lies in the strong AGN region of the WHAN diagram (Fig. \ref{fig:IC1947} bottom panel), which confirms that IC1947 harbours an accreting black hole.

\begin{figure}
    \centering
\includegraphics[width=0.45\textwidth]{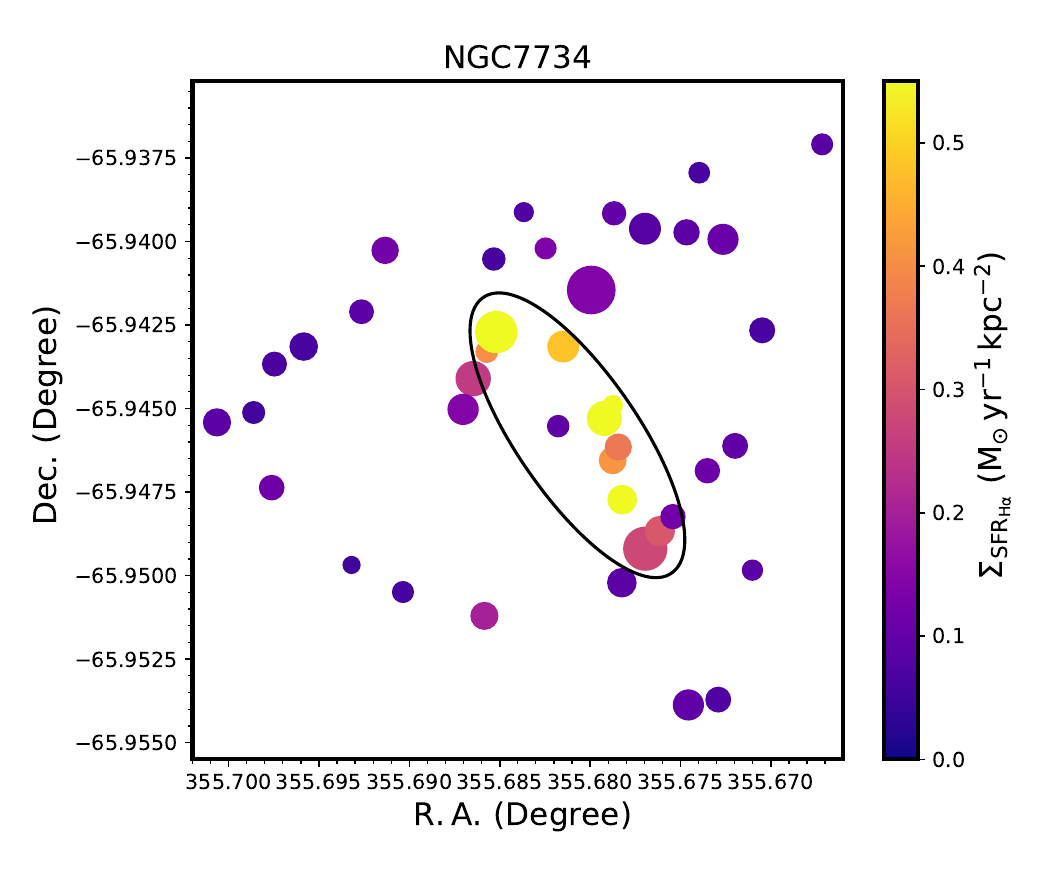}
    \caption{The distribution of detected SFCs in NGC7734 H$\alpha$ map. The SFCs are coloured according to their $\Sigma_{SFR(H\alpha)}$ and scaled according to size. The $\Sigma_{SFR(H\alpha)}$ is higher along and at the bar edges. The black ellipse shows the position of the bar.}
    \label{fig:NGC7734_color_sfr}
\end{figure}

\begin{figure}
    \centering
    \includegraphics[width=0.45\textwidth]{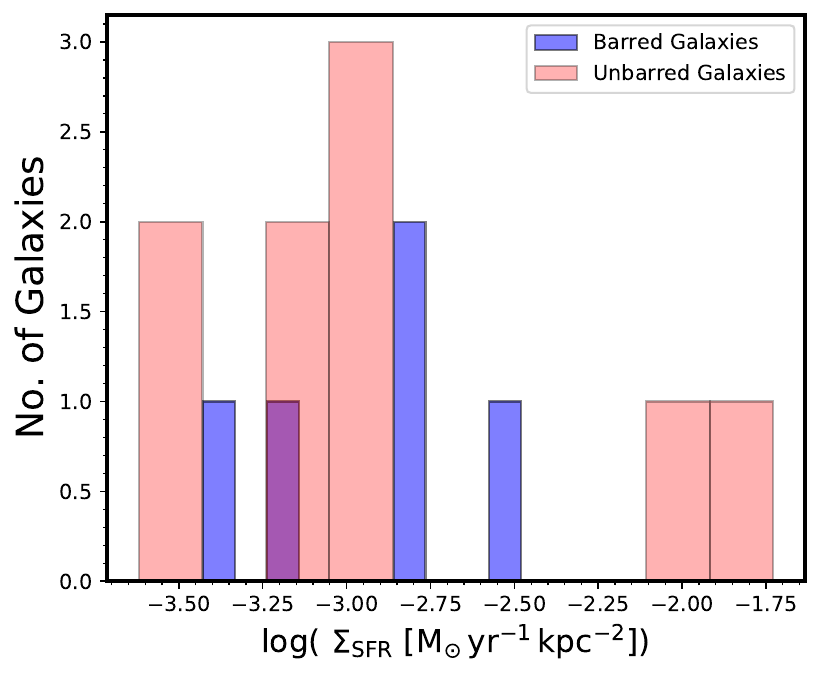}
    \caption{Histograms compare barred and unbarred galaxies in the sample.}
    \label{fig:sfr_bar_unbar}
\end{figure}

\subsection{Effect of the bar on star formation}
There have been several studies on how bars play a role in star formation; however, the conclusions are not clear. Some studies suggest that bars are the channel of gas inflow towards the centre and thus can efficiently drive gas inside corotation to the circum-nuclear region of galaxies. The gas gets compressed and can trigger nuclear star formation \citep{Combes1985A&A...150..327C,Devereux1987ApJ...323...91D,Hummel1990A&A...236..333H}. However, some studies suggest that bars can induce shear and shocks in gas \citep{das.jog.1995}, which leads to turbulence and inhibits star formation along the bar. The star formation then appears quenched \citep{Haywood2016A&A...589A..66H,Khoperskov2018A&A...609A..60K}. Thus, it is important to study the effect of the bar on star formation as there can be different effects, many of which are based on the bar strength. The galaxies NGC7733, NGC7734, NGC646NED01, NGC1356 and ESO235-IG023 have bars in the sample. NGC7733, NGC7734, and ESO235-IG023 NED01 have strong bars, while NGC646NED01 and NGC1356 have intermediate bars. NGC7734 and NGC646NED01 show SFCs along the bar, while NGC7733 shows a few SFCs along the bar and a few at the edges of the bar. We compared the star formation of barred galaxies with those that do not have bars. Fig. \ref{fig:sfr_bar_unbar}
shows the histogram of barred and unbarred galaxies.
The barred and unbarred galaxies in our sample do not show any difference in $\Sigma_{SFR}$; however, the galaxies with bars show higher $\Sigma_{SFR}$ for the SFCs which are located at the bar ends and along the bar compared to the SFCs in the disk (Fig. \ref{fig:NGC7734_color_sfr}). However, a larger sample is required to assess the effect of bars on the distribution of star formation, especially SFCs in the inner disks of galaxies and around bars.

\section{Summary and Conclusions}
In this study, we have presented the properties of eight southern interacting systems. The observations have been carried out using the UVIT and IRSF. We also used MUSE archival data. The good spatial resolution of UVIT was important for detecting a large number of SFCs in the galaxy disks.
\begin{enumerate}
    \item The galaxies in our sample are gas-rich and show ongoing, massive star formation in their tidal tails and in the disk region. Many of the galaxies have star-forming regions associated with the tidal arms.
    \item The surface densities of the SFR $\Sigma_{SFR}$ do not show any correlation with the projected distance between the interacting pair, indicating that the SFR is more affected by local conditions.
    \item The galaxies in the sample show currently active episodes of star formation in their SFH. The ongoing star formation is also evident from FUV and H$\alpha$. 
    \item The galaxies have higher metallicity along the rings and arms, which are the sites of bright SFCs
    \item IC5250B shows a gas outflow up to around five kpc. The disk of both galaxies does not show star formation: this is a dry merger. NGC7733N also shows outflow around eight kpc.
    \item IC1947 hosts an AGN, confirmed using the BPT and WHAN diagrams.
    \item Bars act as a channel of gas inflow, and gas can get compressed and form stars along and at the edges of the bar. The SFCs at the edges of the bar show enhancement in $\Sigma_{SFR}$.
\end{enumerate}
 
\section*{Acknowledgements}
We thank the anonymous referee for the thoughtful review, which improved the impact and clarity of this work. This paper has used the data from the UVIT, which is part of the \textit{AstroSat} mission of the Indian Space Research Organization (ISRO), archived at the Indian Space Science Data Center (ISSDC). This publication has also used NIR data from IRSF at SAAO. This paper has also used the observations collected at the European Southern Observatory under ESO programme 0103.A-0637, run B. This research has also used data from DECaLS at CTIO. This publication has used the NASA/IPAC Extragalactic Database (NED), operated by the Jet Propulsion Laboratory, California Institute of Technology, under contract with the National Aeronautics and Space Administration. MD acknowledges the support of the Science and Engineering Research Board (SERB) MATRICS grant MTR/2020/000266 for this research. JY acknowledges the support of Department of Science for the Science and Engineering Research Board Overseas Visiting Doctoral Fellowship.

\section*{Data Availability}
The UVIT data used in this paper are publicly available at 
\url{https://astrobrowse.issdc.gov.in/astro_archive/archive/Home.jsp}. The Galex data used in the paper are publicly available at \url{https://mast.stsci.edu/portal/Mashup/Clients/Mast/Portal.html}. The MUSE data used in the paper is available at \url{http://archive.eso.org/scienceportal/home}.




\bibliographystyle{mnras}
\bibliography{mnras} 








\bsp	
\label{lastpage}
\end{document}